\def\mezzo#1{\bigskip\noindent{\sl #1}\bigskip}

\def\unredoffs{} \def\redoffs{\voffset=-.40truein\hoffset=-.40truein}
\def\speclscape{}

\newbox\leftpage \newdimen\fullhsize \newdimen\hstitle \newdimen\hsbody
\tolerance=1000\hfuzz=2pt

\catcode`\@=11
\def\bigans{b }

\def\answ{b }

\ifx\answ\bigans\message{(This will come out unreduced.}
\magnification=1200\unredoffs\baselineskip=16pt plus 2pt minus 1pt
\hsbody=\hsize \hstitle=\hsize

\else\message{(This will be reduced.} \let\l@r=L
\magnification=1000\baselineskip=16pt plus 2pt minus 1pt
\vsize=7truein \redoffs
\hstitle=8truein\hsbody=4.75truein\fullhsize=10truein\hsize=\hsbody
\output={\ifnum\pageno=0

   \shipout\vbox{\speclscape{\hsize\fullhsize\makeheadline}
     \hbox to \fullhsize{\hfill\pagebody\hfill}}\advancepageno
   \else
   \almostshipout{\leftline{\vbox{\pagebody\makefootline}}}\advancepageno
   \fi}
\def\almostshipout#1{\if L\l@r \count1=1 \message{[\the\count0.\the\count1]}
       \global\setbox\leftpage=#1 \global\let\l@r=R
  \else \count1=2
   \shipout\vbox{\speclscape{\hsize\fullhsize\makeheadline}
       \hbox to\fullhsize{\box\leftpage\hfil#1}}  \global\let\l@r=L\fi}
\fi

\newcount\yearltd\yearltd=\year\advance\yearltd by -1900

\def\Title#1#2{\nopagenumbers\abstractfont\hsize=\hstitle\rightline{#1}%
\vskip 1in\centerline{\titlefont #2}\abstractfont\vskip
.5in\pageno=0}
\def\Date#1{\vfill\leftline{#1}\tenpoint\supereject\global\hsize=\hsbody%
\footline={\hss\tenrm\folio\hss}}

\def\draftmode{\message{ DRAFTMODE }\def\draftdate{{\rm preliminary draft:
\number\month/\number\day/\number\yearltd\ \ \hourmin}}%
\headline={\hfil\draftdate}\writelabels\baselineskip=20pt plus 2pt
minus 2pt
  {\count255=\time\divide\count255 by 60 \xdef\hourmin{\number\count255}
   \multiply\count255 by-60\advance\count255 by\time
   \xdef\hourmin{\hourmin:\ifnum\count255<10 0\fi\the\count255}}}

\def\nolabels{\def\wrlabeL##1{}\def\eqlabeL##1{}\def\reflabeL##1{}}
\def\writelabels{\def\wrlabeL##1{\leavevmode\vadjust{\rlap{\smash%
{\line{{\escapechar=` \hfill\rlap{\sevenrm\hskip.03in\string##1}}}}}}}%
\def\eqlabeL##1{{\escapechar-1\rlap{\sevenrm\hskip.05in\string##1}}}%
\def\reflabeL##1{\noexpand\llap{\noexpand\sevenrm\string\string\string##1}}}
\nolabels

\global\newcount\secno \global\secno=0 \global\newcount\meqno
\global\meqno=1
\def\newsec#1{\global\advance\secno by1\message{(\the\secno. #1)}

\global\subsecno=0\eqnres@t\noindent{\bf\the\secno. #1}
\writetoca{{\secsym} {#1}}\par\nobreak\medskip\nobreak}
\def\eqnres@t{\xdef\secsym{\the\secno.}\global\meqno=1\bigbreak\bigskip}
\def\sequentialequations{\def\eqnres@t{\bigbreak}}
\xdef\secsym{}
\global\newcount\subsecno \global\subsecno=0
\def\subsec#1{\global\advance\subsecno by1\message{(\secsym\the\subsecno. #1)}
\ifnum\lastpenalty>9000\else\bigbreak\fi
\noindent{\it\secsym\the\subsecno. #1}\writetoca{\string\quad
{\secsym\the\subsecno.} {#1}}\par\nobreak\medskip\nobreak}
\def\appendix#1#2{\global\meqno=1\global\subsecno=0\xdef\secsym{\hbox{#1.}}
\bigbreak\bigskip\noindent{\bf Appendix #1. #2}\message{(#1. #2)}
\writetoca{Appendix {#1.} {#2}}\par\nobreak\medskip\nobreak}

\def\eqnn#1{\xdef #1{(\secsym\the\meqno)}\writedef{#1\leftbracket#1}%
\global\advance\meqno by1\wrlabeL#1}
\def\eqna#1{\xdef #1##1{\hbox{$(\secsym\the\meqno##1)$}}
\writedef{#1\numbersign1\leftbracket#1{\numbersign1}}%
\global\advance\meqno by1\wrlabeL{#1$\{\}$}}
\def\eqn#1#2{\xdef #1{(\secsym\the\meqno)}\writedef{#1\leftbracket#1}%
\global\advance\meqno by1$$#2\eqno#1\eqlabeL#1$$}

\newskip\footskip\footskip14pt plus 1pt minus 1pt

\def\footnotefont{\ninepoint}\def\f@t#1{\footnotefont #1\@foot}
\def\f@@t{\baselineskip\footskip\bgroup\footnotefont\aftergroup\@foot\let\next}
\setbox\strutbox=\hbox{\vrule height9.5pt depth4.5pt width0pt}
\global\newcount\ftno \global\ftno=0
\def\foot{\global\advance\ftno by1\footnote{$^{\the\ftno}$}}

\newwrite\ftfile
\def\footend{\def\foot{\global\advance\ftno by1\chardef\wfile=\ftfile
$^{\the\ftno}$\ifnum\ftno=1\immediate\openout\ftfile=foots.tmp\fi%
\immediate\write\ftfile{\noexpand\smallskip%
\noexpand\item{f\the\ftno:\ }\pctsign}\findarg}%
\def\footatend{\vfill\eject\immediate\closeout\ftfile{\parindent=20pt
\centerline{\bf Footnotes}\nobreak\bigskip\input foots.tmp }}}
\def\footatend{}

\global\newcount\refno \global\refno=1
\newwrite\rfile
\def\ref{[\the\refno]\nref}
\def\nref#1{\xdef#1{[\the\refno]}\writedef{#1\leftbracket#1}%
\ifnum\refno=1\immediate\openout\rfile=refs.tmp\fi
\global\advance\refno by1\chardef\wfile=\rfile\immediate
\write\rfile{\noexpand\item{#1\
}\reflabeL{#1\hskip.31in}\pctsign}\findarg}

\def\findarg#1#{\begingroup\obeylines\newlinechar=`\^^M\pass@rg}
{\obeylines\gdef\pass@rg#1{\writ@line\relax #1^^M\hbox{}^^M}%
\gdef\writ@line#1^^M{\expandafter\toks0\expandafter{\striprel@x #1}%
\edef\next{\the\toks0}\ifx\next\em@rk\let\next=\endgroup\else\ifx\next\empty%
\else\immediate\write\wfile{\the\toks0}\fi\let\next=\writ@line\fi\next\relax}}
\def\striprel@x#1{} \def\em@rk{\hbox{}}
\def\lref{\begingroup\obeylines\lr@f}
\def\lr@f#1#2{\gdef#1{\ref#1{#2}}\endgroup\unskip}
\def\semi{;\hfil\break}
\def\addref#1{\immediate\write\rfile{\noexpand\item{}#1}}

\def\startrefs#1{\immediate\openout\rfile=refs.tmp\refno=#1}
\def\xref{\expandafter\xr@f}\def\xr@f[#1]{#1}
\def\refs#1{\count255=1[\r@fs #1{\hbox{}}]}
\def\r@fs#1{\ifx\und@fined#1\message{reflabel \string#1 is undefined.}%
\nref#1{need to supply reference \string#1.}\fi%
\vphantom{\hphantom{#1}}\edef\next{#1}\ifx\next\em@rk\def\next{}%
\else\ifx\next#1\ifodd\count255\relax\xref#1\count255=0\fi%
\else#1\count255=1\fi\let\next=\r@fs\fi\next}

\newwrite\ffile\global\newcount\figno \global\figno=1
\def\fig{fig.~\the\figno\nfig}
\def\nfig#1{\xdef#1{fig.~\the\figno}%
\writedef{#1\leftbracket fig.\noexpand~\the\figno}%
\ifnum\figno=1\immediate\openout\ffile=figs.tmp\fi\chardef\wfile=\ffile%
\immediate\write\ffile{\noexpand\medskip\noexpand\item{Fig.\
\the\figno. }
\reflabeL{#1\hskip.55in}\pctsign}\global\advance\figno
by1\findarg}
\def\xfig{\expandafter\xf@g}\def\xf@g fig.\penalty\@M\ {}
\def\figs#1{figs.~\f@gs #1{\hbox{}}}
\def\f@gs#1{\edef\next{#1}\ifx\next\em@rk\def\next{}\else
\ifx\next#1\xfig #1\else#1\fi\let\next=\f@gs\fi\next}
\newwrite\lfile
{\escapechar-1\xdef\pctsign{\string\%}\xdef\leftbracket{\string\{}
\xdef\rightbracket{\string\}}\xdef\numbersign{\string\#}}

\def\writestop{\def\writestoppt{\immediate\write\lfile{\string\pageno%
\the\pageno\string\startrefs\leftbracket\the\refno\rightbracket%
\string\def\string\secsym\leftbracket\secsym\rightbracket%
\string\secno\the\secno\string\meqno\the\meqno}\immediate\closeout\lfile}}
\def\writestoppt{}\def\writedef#1{}
\def\seclab#1{\xdef #1{\the\secno}\writedef{#1\leftbracket#1}\wrlabeL{#1=#1}}
\def\subseclab#1{\xdef #1{\secsym\the\subsecno}%
\writedef{#1\leftbracket#1}\wrlabeL{#1=#1}}
\newwrite\tfile \def\writetoca#1{}
\def\leaderfill{\leaders\hbox to 1em{\hss.\hss}\hfill}

\def\writetoc{\immediate\openout\tfile=SWN.tmp
    \def\writetoca##1{{\edef\next{\write\tfile{\noindent ##1
    \string\leaderfill {\noexpand\number\pageno} \par}}\next}}}


%
\catcode`\@=12 
%
\edef\tfontsize{\ifx\answ\bigans scaled\magstep3\else
scaled\magstep4\fi} \font\titlerm=cmr10 \tfontsize
\font\titlerms=cmr7 \tfontsize \font\titlermss=cmr5 \tfontsize
\font\titlei=cmmi10 \tfontsize \font\titleis=cmmi7 \tfontsize
\font\titleiss=cmmi5 \tfontsize \font\titlesy=cmsy10 \tfontsize
\font\titlesys=cmsy7 \tfontsize \font\titlesyss=cmsy5 \tfontsize
\font\titleit=cmti10 \tfontsize \skewchar\titlei='177
\skewchar\titleis='177 \skewchar\titleiss='177
\skewchar\titlesy='60 \skewchar\titlesys='60
\skewchar\titlesyss='60
\def\titlefont{\def\rm{\fam0\titlerm}
\textfont0=\titlerm \scriptfont0=\titlerms
\scriptscriptfont0=\titlermss \textfont1=\titlei
\scriptfont1=\titleis \scriptscriptfont1=\titleiss
\textfont2=\titlesy \scriptfont2=\titlesys
\scriptscriptfont2=\titlesyss \textfont\itfam=\titleit
\def\it{\fam\itfam\titleit}\rm}
 \ifx\answ\bigans\else scaled\magstep1\fi
\ifx\answ\bigans\def\abstractfont{\tenpoint}\else
\font\abssl=cmsl10 scaled \magstep1 \font\absrm=cmr10
scaled\magstep1 \font\absrms=cmr7 scaled\magstep1
\font\absrmss=cmr5 scaled\magstep1 \font\absi=cmmi10
scaled\magstep1 \font\absis=cmmi7 scaled\magstep1
\font\absiss=cmmi5 scaled\magstep1 \font\abssy=cmsy10
scaled\magstep1 \font\abssys=cmsy7 scaled\magstep1
\font\abssyss=cmsy5 scaled\magstep1 \font\absbf=cmbx10
scaled\magstep1 \skewchar\absi='177 \skewchar\absis='177
\skewchar\absiss='177 \skewchar\abssy='60 \skewchar\abssys='60
\skewchar\abssyss='60
\def\abstractfont{\def\rm{\fam0\absrm}
\textfont0=\absrm \scriptfont0=\absrms \scriptscriptfont0=\absrmss
\textfont1=\absi \scriptfont1=\absis \scriptscriptfont1=\absiss
\textfont2=\abssy \scriptfont2=\abssys \scriptscriptfont2=\abssyss
\textfont\itfam=\bigit \def\it{\fam\itfam\bigit}\def\footnotefont{\tenpoint}%
\textfont\slfam=\abssl \def\sl{\fam\slfam\abssl}%
\textfont\bffam=\absbf \def\bf{\fam\bffam\absbf}\rm}\fi
\def\tenpoint{\def\rm{\fam0\tenrm}
\textfont0=\tenrm \scriptfont0=\sevenrm \scriptscriptfont0=\fiverm
\textfont1=\teni  \scriptfont1=\seveni  \scriptscriptfont1=\fivei
\textfont2=\tensy \scriptfont2=\sevensy \scriptscriptfont2=\fivesy
\textfont\itfam=\tenit \def\it{\fam\itfam\tenit}\def\footnotefont{\ninepoint}%
\textfont\bffam=\tenbf
\def\bf{\fam\bffam\tenbf}\def\sl{\fam\slfam\tensl}\rm}
\font\ninerm=cmr9 \font\sixrm=cmr6 \font\ninei=cmmi9
\font\sixi=cmmi6 \font\ninesy=cmsy9 \font\sixsy=cmsy6
\font\ninebf=cmbx9 \font\nineit=cmti9 \font\ninesl=cmsl9
\skewchar\ninei='177 \skewchar\sixi='177 \skewchar\ninesy='60
\skewchar\sixsy='60
\def\ninepoint{\def\rm{\fam0\ninerm}
\textfont0=\ninerm \scriptfont0=\sixrm \scriptscriptfont0=\fiverm
\textfont1=\ninei \scriptfont1=\sixi \scriptscriptfont1=\fivei
\textfont2=\ninesy \scriptfont2=\sixsy \scriptscriptfont2=\fivesy
\textfont\itfam=\ninei \def\it{\fam\itfam\nineit}\def\sl{\fam\slfam\ninesl}%
\textfont\bffam=\ninebf \def\bf{\fam\bffam\ninebf}\rm}
%
%

\hyphenation{anom-aly anom-alies coun-ter-term coun-ter-terms}
\def\inv{^{\raise.15ex\hbox{${\scriptscriptstyle -}$}\kern-.05em 1}}

\def\Dsl{\,\raise.15ex\hbox{/}\mkern-13.5mu D} 
\def\dsl{\raise.15ex\hbox{/}\kern-.57em\partial}

 \def\Tr{{\rm Tr}}
\font\bigit=cmti10 scaled \magstep1
\def\lspace{\ifx\answ\bigans{}\else\qquad\fi}
\def\lbspace{\ifx\answ\bigans{}\else\hskip-.2in\fi} 
\def\boxeqn#1{\vcenter{\vbox{\hrule\hbox{\vrule\kern3pt\vbox{\kern3pt
     \hbox{${\displaystyle #1}$}\kern3pt}\kern3pt\vrule}\hrule}}}
\def\mbox#1#2{\vcenter{\hrule \hbox{\vrule height#2in
         \kern#1in \vrule} \hrule}}  

\def\tilde{\widetilde}
\def\bar{\overline}
\def\hat{\widehat}

 \def\CO{{\cal O}}
\def\CA{{\cal A}}  \def\CF{{\cal F}} \def\CG{{\cal G}}
 \def\CH{{\cal H}} \def\CI{{\cal I}} 
 \def\CR{{\cal R}} \def\CD{{\cal D}} 
\def\e#1{{\rm e}^{^{\textstyle#1}}}

\def\darr#1{\raise1.5ex\hbox{$\leftrightarrow$}\mkern-16.5mu #1}

\def\half{{\textstyle{1\over2}}}

\def\roughly#1{\raise.3ex\hbox{$#1$\kern-.75em\lower1ex\hbox{$\sim$}}}

\def\np#1#2#3{Nucl. Phys. {\bf B#1} (#2) #3}
\def\pl#1#2#3{Phys. Lett. {\bf #1B} (#2) #3}

\def\anp#1#2#3{Ann. Phys. {\bf #1} (#2) #3}

\def\cmp#1#2#3{Comm. Math. Phys. {\bf #1} (#2) #3}

\def\jhep#1#2#3{JHEP {\bf#1}(#2) #3}

\def\atmp#1#2#3{Adv.~Theor.~Math.~Phys.{\bf #1} (#2) #3}

\def\IB{\relax\hbox{$\inbar\kern-.3em{\rm B}$}}
\def\IC{\relax\hbox{$\inbar\kern-.3em{\rm C}$}}
\def\ID{\relax\hbox{$\inbar\kern-.3em{\rm D}$}}
\def\IE{\relax\hbox{$\inbar\kern-.3em{\rm E}$}}
\def\IF{\relax\hbox{$\inbar\kern-.3em{\rm F}$}}
\def\IG{\relax\hbox{$\inbar\kern-.3em{\rm G}$}}
\def\IGa{\relax\hbox{${\rm I}\kern-.18em\Gamma$}}
\def\IH{\relax{\rm I\kern-.18em H}}
\def\IK{\relax{\rm I\kern-.18em K}}
\def\IL{\relax{\rm I\kern-.18em L}}
\def\IP{\relax{\rm I\kern-.18em P}}
\def\IR{\relax{\rm I\kern-.18em R}}
\def\IZ{\relax\ifmmode\mathchoice{
\hbox{\cmss Z\kern-.4em Z}}{\hbox{\cmss Z\kern-.4em Z}}
{\lower.9pt\hbox{\cmsss Z\kern-.4em Z}} {\lower1.2pt\hbox{\cmsss
Z\kern-.4em Z}} \else{\cmss Z\kern-.4em Z}\fi}
\def\II{\relax{\rm I\kern-.18em I}}


\def\CA{{\cal A}}

\def\CD{{\cal D}}
\def\CE{{\cal E}}
\def\CF{{\cal F}}
\def\CG{{\cal G}}
\def\CH{{\cal H}}
\def\CI{{\cal I}}

\def\CM{{\cal M}}
\def\CN{{\cal N}}
\def\CO{{\cal O}}
\def\CP{{\cal P}}

\def\CR{{\cal R}}
\def\CS{{\cal S}}

\def\CV{{\cal V}}
\def\CW{{\cal W}}
\def\CX{{\cal X}}
\def\CY{{\cal Y}}

\def\p{\partial}
\def\pb{\bar{\partial}}
\def\dir{{\CD}\hskip -6pt \slash \hskip 5pt}

\def\ib{\bar{i}}
\def\jb{\bar{j}}

\def\wb{\bar{w}}

\def\zb{\bar{z}}

\def\Tr{{\rm Tr}}

\def\sdtimes{\mathbin{\hbox{\hskip2pt\vrule height 4.1pt depth -.3pt
width.25pt\hskip-2pt$\times$}}}

\def\Det{{\rm Det}}


\def\inbar{\,\vrule height1.5ex width.4pt depth0pt}

\font\cmss=cmss10 \font\cmsss=cmss10 at 7pt

\def\sdtimes{\mathbin{\hbox{\hskip2pt\vrule height 4.1pt
depth -.3pt width .25pt\hskip-2pt$\times$}}}
\def\a{{\alpha}}

\def\b{{\beta}}
\def\d{{\delta}}

\def\e{{\epsilon}}
\def\z{{\zeta}}
\def\ve{{\varepsilon}}
\def\vf{{\varphi}}
\def\m{{\mu}}
\def\n{{\nu}}
\def\u{{\Upsilon}}
\def\l{{\lambda}}
\def\s{{\sigma}}
\def\t{{\theta}}

\def\o{{\omega}}
\def\nc{noncommutative\ }

\def\k{{\kappa}}
\def\bA{{\bf A}}

\def\boxit#1{\vbox{\hrule\hbox{\vrule\kern8pt
\vbox{\hbox{\kern8pt}\hbox{\vbox{#1}}\hbox{\kern8pt}}
\kern8pt\vrule}\hrule}}
\def\mathboxit#1{\vbox{\hrule\hbox{\vrule\kern8pt\vbox{\kern8pt
\hbox{$\displaystyle #1$}\kern8pt}\kern8pt\vrule}\hrule}}


\chardef\tempcat=\the\catcode`\@ \catcode`\@=11
\def\cyracc{\def\u##1{\if \i##1\accent"24 i%
    \else \accent"24 ##1\fi }}
\newfam\cyrfam
\font\tencyr=wncyr10
\def\cyr{\fam\cyrfam\tencyr\cyracc}


\def\CA{{\cal A}} 
\def\CF{{\cal F}} \def\CG{{\cal G}} 
\def\CH{{\cal H}} \def\CI{{\cal I}} 
 \def\CO{{\cal O}}
\def\CR{{\cal R}} \def\CD{{\cal D}}


\def\e#1{{\rm e}^{^{\textstyle#1}}}

\def\darr#1{\raise1.5ex\hbox{$\leftrightarrow$}\mkern-16.5mu #1}

\def\half{{\textstyle{1\over2}}} 

\def\roughly#1{\raise.3ex\hbox{$#1$\kern-.75em\lower1ex\hbox{$\sim$}}}


\def\lref{\begingroup\obeylines\lr@f}
\def\lr@f#1#2{\gdef#1{\ref#1{#2}}\endgroup\unskip}

\def\np#1#2#3{Nucl. Phys. {\bf B#1} (#2) #3}
\def\pl#1#2#3{Phys.Lett. {\bf #1B} (#2) #3}

\def\anp#1#2#3{Ann. Phys. {\bf #1} (#2) #3}

\def\cmp#1#2#3{Comm.Math. Phys. {\bf #1} (#2) #3}
 
\def\jhep#1#2#3{JHEP {\bf#1}(#2) #3}

\def\atmp#1#2#3{Adv.~Theor.~Math.~Phys.{\bf #1} (#2) #3}


\def\IB{\relax\hbox{$\inbar\kern-.3em{\rm B}$}}
\def\IC{\relax\hbox{$\inbar\kern-.3em{\rm C}$}}
\def\ID{\relax\hbox{$\inbar\kern-.3em{\rm D}$}}
\def\IE{\relax\hbox{$\inbar\kern-.3em{\rm E}$}}
\def\IF{\relax\hbox{$\inbar\kern-.3em{\rm F}$}}
\def\IG{\relax\hbox{$\inbar\kern-.3em{\rm G}$}}
\def\IGa{\relax\hbox{${\rm I}\kern-.18em\Gamma$}}
\def\IH{\relax{\rm I\kern-.18em H}} \def\IK{\relax{\rm
I\kern-.18em K}} \def\IL{\relax{\rm I\kern-.18em L}}
\def\IP{\relax{\rm I\kern-.18em P}} \def\IR{\relax{\rm
I\kern-.18em R}} \def\IZ{\relax\ifmmode\mathchoice{ \hbox{\cmss
Z\kern-.4em Z}}{\hbox{\cmss Z\kern-.4em Z}}
{\lower.9pt\hbox{\cmsss Z\kern-.4em Z}} {\lower1.2pt\hbox{\cmsss
Z\kern-.4em Z}} \else{\cmss Z\kern-.4em Z}\fi} \def\II{\relax{\rm
I\kern-.18em I}}

  \def\CD{{\cal D}}
\def\CE{{\cal E}} \def\CF{{\cal F}} \def\CG{{\cal G}}
\def\CH{{\cal H}} \def\CI{{\cal I}} 
  \def\CM{{\cal M}}
\def\CN{{\cal N}} \def\CO{{\cal O}} \def\CP{{\cal P}}
 \def\CR{{\cal R}} \def\CS{{\cal S}}
  \def\CV{{\cal V}}
\def\CW{{\cal W}} \def\CX{{\cal X}} \def\CY{{\cal Y}}


\def\p{\partial} \def\pb{\bar{\partial}}
\def\dir{{\CD}\hskip -6pt \slash \hskip 5pt}


\def\ib{\bar{i}}
\def\jb{\bar{j}}  
  
\def\wb{\bar{w}}  
\def\zb{\bar{z}}


  \def\Tr{{\rm Tr}}
  
 \def\sdtimes{\mathbin{\hbox{\hskip2pt\vrule height
4.1pt depth -.3pt width.25pt\hskip-2pt$\times$}}}  \def\Det{{\rm Det}}


\def\inbar{\,\vrule height1.5ex width.4pt depth0pt}
\font\cmss=cmss10 \font\cmsss=cmss10 at 7pt
\def\sdtimes{\mathbin{\hbox{\hskip2pt\vrule height 4.1pt depth
-.3pt width .25pt\hskip-2pt$\times$}}}


\def\a{{\alpha}} 
\def\b{{\beta}} \def\d{{\delta}}  
\def\e{{\epsilon}} \def\z{{\zeta}} \def\ve{{\varepsilon}}
\def\vf{{\varphi}} \def\m{{\mu}} \def\n{{\nu}} \def\u{{\Upsilon}}
\def\l{{\lambda}} \def\s{{\sigma}} \def\t{{\theta}}
 \def\o{{\omega}} \def\nc{noncommutative\ }


\def\ba{{\bf a}}  \def\bg{{\bf g}} 
\def\bG{{\bf G}} \def\bT{{\bf T}} \def\bX{{\bf X}}
\def\bR{{\bf R}} \def\bZ{{\bf Z}} \def\bC{{\bf C}} \def\bP{{\bf P}}
\def\bS{{\bf S}}   \def\vf{{\bf f}}
  \def\bk{{\bf k}}
  \def\bt{{\bf t}}

\def\bM{{\bf M}}
\def\bI{{\bf I}}
\def\bW{{\bf W}}

 \def\IF{{\bf F}}
\def\boxit#1{\vbox{\hrule\hbox{\vrule\kern8pt
\vbox{\hbox{\kern8pt}\hbox{\vbox{#1}}\hbox{\kern8pt}}
\kern8pt\vrule}\hrule}}
\def\mathboxit#1{\vbox{\hrule\hbox{\vrule\kern8pt\vbox{\kern8pt
\hbox{$\displaystyle #1$}\kern8pt}\kern8pt\vrule}\hrule}}

\lref\tdym{D.~Gross, hep-th/9212149\semi D.~Gross, W.~Taylor,
hep-th/9301068, hep-th/9303046}

\lref\dougcft{M.~Douglas, hep-th/9311130, hep-th/9303159}

\lref\yung{A.~M.~Vershik, ``Hook formulae and related
identities'', {\cyr Zapiski sem. LOMI}, {\bf 172} (1989), 3-20 (in
Russian)\semi S.~V.~Kerov, A.~M.~Vershik, ``Asymptotics of the
Plancherel measure of the symmetric group and the limiting shape
of the Young diagrams'', {\cyr DAN SSSR}, {\bf 233} (1977),
1024-1027 (in Russian)\semi S.~V.~Kerov, ``Random Young
tableaux'', {\cyr Teor. veroyat. i ee primeneniya}, {\bf 3}
(1986), 627-628 (in Russian)}
 \lref\phtran{M.~Douglas, V.~Kazakov, hep-th/9305047}

\lref\dv{R.~Dijkgraaf, C.~Vafa, hep-th/0206255, hep-th/0207106,
hep-th/0208048\semi R.~Dijkgraaf, S.~Gukov, V.~Kazakov, C.~Vafa,
hep-th/0210238\semi R.~Dijkgraaf, M.~Grisaru, C.~Lam, C.~Vafa,
D.~Zanon, hep-th/0211017\semi M.~Aganagic, M.~Marino, A.~Klemm,
C.~Vafa, hep-th/0211098\semi R.~Dijkgraaf, A.~Neitzke, C.~Vafa,
hep-th/0211194} \lref\cdws{F.~Cachazo, M.~Douglas, N.~Seiberg,
E.~Witten, hep-th/0211170}

\lref\estring{J.~A.~Minahan, D.~Nemeschansky, C.~Vafa, N.P.~Warner,
hep-th/9802168\semi
T.~Eguchi, K.~Sakai, hep-th/0203025, hep-th/0211213}

\lref\bulkbndr{V.~Balasubramanian, P.~Kraus, A.~Lawrence, hep-th/9805171}

\lref\frbranes{D.-E.~Diaconescu, M.~Douglas, J.~Gomis, hep-th/9712230}

\lref\gorodentsevleenzon{A.~Klyachko, ``Moduli of vector bundles
and numbers of classes'', Funct. Anal. and Appl. {\bf 25} (1991)
67\semi G.~Ellingsrud, L.~G\"ottsche, alg-geom/9506019\semi
A.~Gorodentsev, M.~Leenson, alg-geom/9604011}
\lref\witdgt{E.~Witten, hep-th/9204083} \lref\ikkt{N.~Ishibashi,
H.~Kawai, Y.~Kitazawa, and A.~Tsuchiya, \np{498}{1997}{467},
hep-th/9612115} \lref\cds{A.~Connes, M.~Douglas, A.~Schwarz,
\jhep{9802}{1998}{003}} \lref\wtnc{E.~Witten, \np{268}{1986}{253}}

\lref\cstw{L.~Baulieu, A.~Losev, N.~Nekrasov, hep-th/9707174}

\lref\klemmzaslow{ A.~Klemm, E.~Zaslow, hep-th/9906046}

\lref\walg{A.~Gerasimov, A.~Levin, A.~Marshakov,
\np{360}{1991}{537}\semi A.~Bilal, I.~Kogan, V.~Fock,
\np{359}{1991}{635}}

\lref\cftorb{A.~Lawrence, N.~Nekrasov, C.~Vafa, hep-th/9803015}
\lref\booksSW{
A.~Marshakov, ``Seiberg-Witten Theory and Integrable Systems,'' {\it
World Scientific, Singapore (1999)}\semi
``Integrability: The Seiberg-Witten and Whitham Equations",
Eds. H.~Braden and I.~Krichever, {\it Gordon and Breach (2000)}.}

\lref\agmav{M.~Aganagic, M.~Mari\~no, C.~Vafa, hep-th/0206164}

\lref\moorewitten{G.~Moore, E.~Witten, hep-th/9709193}

\lref\gopakumarvafa{R.~Gopakumar, C.Vafa, hep-th/9809187,
hep-th/9812127} \lref\mooreunpublished{G.~Moore, unpublished}
\lref\wittenone{E.~Witten, hep-th/9403195} \lref\cg{E.~Corrigan,
P.~Goddard, ``Construction of instanton and monopole solutions and
reciprocity'', \anp {154}{1984}{253}} \lref\opennc{N.~Nekrasov,
hep-th/0010017,
hep-th/0203109}\lref\kly{K.-Y.Kim, B.-H. Lee, H.S. Yang, hep-th/0205010
} \lref\donaldson{S.K.~Donaldson, ``Instantons and Geometric
Invariant Theory", \cmp{93}{1984}{453-460}}
\lref\nakajima{H.~Nakajima, ``Lectures on Hilbert Schemes of
Points on Surfaces''\semi AMS University Lecture Series, 1999,
ISBN 0-8218-1956-9. } \lref\neksch{N.~Nekrasov, A.~S.~Schwarz,
hep-th/9802068, \cmp{198}{1998}{689}} \lref\freck{A.~Losev,
N.~Nekrasov, S.~Shatashvili, hep-th/9908204, hep-th/9911099}
\lref\rkh{N.J.~Hitchin, A.~Karlhede, U.~Lindstrom, and M.~Rocek,
\cmp{108}{1987}{535}} \lref\branek{H.~Braden, N.~Nekrasov,
hep-th/9912019}\lref\kazuyuki{K.~Furuuchi, hep-th/9912047} \lref\wilson{G.~
Wilson, ``Collisions of Calogero-Moser particles and adelic
Grassmannian", Invent. Math. 133 (1998) 1-41.}
\lref\abs{O.~Aharony, M.~Berkooz, N.~Seiberg, hep-th/9712117,
\atmp{2}{1998}{119-153}} \lref\avatars{A.~Losev, G.~Moore,
N.~Nekrasov, S.~Shatashvili, hep-th/9509151}
\lref\abkss{O.~Aharony, M.~Berkooz, S.~Kachru, N.~Seiberg,
E.~Silverstein, hep-th/9707079, \atmp{1}{1998}{148-157}}

\lref\cecotti{S.~Cecotti, L.~Girardello, \pl{110}{1982}{39}}
\lref\smilga{A.~Smilga, Yad.Fiz. {\bf 43} (1986), 215-218}
\lref\sethi{S.~Sethi, M.~Stern, hep-th/9705046}

\lref\witsei{N.~Seiberg, E.~Witten, hep-th/9908142,
\jhep{9909}{1999}{032}} \lref\kkn{V.~Kazakov, I.~Kostov,
N.~Nekrasov, ``D-particles, Matrix Integrals and KP hierarchy'',
\np{557}{1999}{413-442}, hep-th/9810035}
\lref\DHf{J.~J.~Duistermaat, G.J.~Heckman, Invent. Math. {\bf 69}
(1982) 259\semi M.~Atiyah, R.~Bott,  Topology {\bf 23} No 1 (1984)
1} \lref\tdgt{M.~Atiyah, R.~Bott,  Phil. Trans. Roy. Soc. London
{\bf A 308} (1982), 524-615\semi E.~Witten, hep-th/9204083\semi
S.~Cordes, G.~Moore, S.~Rangoolam, hep-th/9411210}
\lref\atiyahsegal{M.~Atiyah, G.~Segal, Ann. of Math. {\bf 87}
(1968) 531} \lref\bott{R.~Bott, J.~Diff.~Geom. {\bf 4} (1967) 311}
\lref\torusaction{G.~Ellingsrud, S.A.Stromme, Invent. Math. {\bf
87} (1987) 343-352\semi L.~G\"ottche, Math. A.. {\bf 286} (1990)
193-207} \lref\gravilit{M.~Bershadsky, S.~Cecotti, H.~Ooguri,
C.~Vafa, \cmp{165}{1994}{311}, \np{405}{1993}{279}\semi
I.~Antoniadis, E.~Gava, K.S.~Narain, T.~R.~Taylor,
\np{413}{1994}{162}, \np{455}{1995}{109}} \lref\calculus{N.~Dorey,
T.~J.~Hollowood, V.~V.~Khoze, M.~P.~Mattis, hep-th/0206063}
\lref\instmeasures{N.~Dorey, V.V.~Khoze, M.P.~Mattis,
hep-th/9706007, hep-th/9708036} \lref\twoinst{N.~Dorey,
V.V.~Khoze, M.P.~Mattis, hep-th/9607066} \lref\vafaengine{S.~Katz,
A.~Klemm, C.~Vafa, hep-th/9609239} \lref\connes{A.~Connes,
``Noncommutative geometry'', Academic Press (1994)}
\lref\macdonald{I.~Macdonald, ``Symmetric functions and Hall
polynomials'', Clarendon Press, Oxford, 1979}
\lref\nikfive{N.~Nekrasov, hep-th/9609219 \semi A.~Lawrence,
N.~Nekrasov, hep-th/9706025}

\lref\fivedim{A.~Marshakov, A.~Mironov, hep-th/9711156\semi
H.~Braden, A.~Marshakov, A.~Mironov, A.~Morozov,
hep-th/9812078,
hep-th/9902205\semi
T.~Eguchi, H.~Kanno, hep-th/0005008\semi
H.~Braden, A.~Marshakov,
hep-th/0009060\semi
 H. Braden, A. Gorsky, A. Odesskii, V. Rubtsov, hep-th/01111066\semi
C.Csaki, J.Erlich, V.V.Khoze, E.Poppitz, Y.Shadmi, Y.Shirman, hep-th/0110188\semi
T.~Hollowood, hep-th/0302165}

\lref\seibergfive{N.~Seiberg, hep-th/9608111}
\lref\ganor{O.~Ganor, hep-th/9607092, hep-th/9608108}

\lref\equivsheaf{A.~Knutsen, E.~Sharpe, hep-th/9804075}

\lref\bcov{ M.~Bershadsky, S.~Cecotti, H.~Ooguri, C.~Vafa, hep-th/9309140}

\lref\op{A.~Okounkov, R.~Pandharipande, math.AG/0207233,
math.AG/0204305} \lref\prtoda{T.~Eguchi, K.~Hori, C.-S.~Xiong,
hep-th/9605225\semi T.~Eguchi, S.~Yang, hep-th/9407134\semi
T.~Eguchi, H.~Kanno, hep-th/9404056}

\lref\givental{A.~Givental, alg-geom/9603021}

\lref\maxim{M.~Kontsevich, hep-th/9405035}
\lref\whitham{A.~Gorsky, A.~Marshakov, A.~Mironov, A.~Morozov,
Nucl.Phys. {\bf B527} (1998) 690-716,
hep-th/9802007} \lref\kricheverwhitham{I.~Krichever,
hep-th/9205110, \cmp{143}{1992}{415}}

\lref\sw{N.~Seiberg, E.~Witten, hep-th/9407087, hep-th/9408099}

\lref\swsol{A.~Klemm, W.~Lerche, S.~Theisen, S.~Yankielowicz,
hep-th/9411048 \semi P.~Argyres, A.~Faraggi, hep-th/9411057\semi
A.~Hanany, Y.~Oz, hep-th/9505074} \lref\hollowood{T.~Hollowood,
hep-th/0201075, hep-th/0202197} \lref\nsvz{V.~Novikov, M.~Shifman,
A.~Vainshtein, V.~Zakharov, \pl{217}{1989}{103}}
\lref\seibergone{N.~Seiberg, \pl{206}{1988}{75}}
\lref\ihiggs{G.~Moore, N.~Nekrasov,
S.~Shatashvili, hep-th/9712241, hep-th/9803265 } \lref\potsdam{W.~Krauth,
H.~Nicolai, M.~Staudacher, hep-th/9803117} \lref\kirwan{F.~Kirwan,
``Cohomology of quotients in symplectic and algebraic geometry'',
Mathematical Notes, Princeton Univ. Press, 1985}
\lref\wittfivebrane{E.~Witten, hep-th/9610234}
\lref\issues{A.~Losev, N.~Nekrasov, S.~Shatashvili,
hep-th/9711108, hep-th/9801061}

\lref\adhm{M.~Atiyah, V.~Drinfeld, N.~Hitchin, Yu.~Manin, Phys.
Lett. {\bf 65A} (1978) 185}

\lref\vafaoo{C.~Vafa, hep-th/0008142}

\lref\seiberghol{N.~Seiberg, hep-th/9408013}

 \lref\warner{A.~Klemm, W.~Lerche, P.~Mayr,
C.~Vafa, N.~Warner, hep-th/9604034}

\lref\wittensolution{E.~Witten, hep-th/9703166}

\lref\witbound{E.~Witten, hep-th/9510153}

\lref\twists{E.~Witten, hep-th/9304026 \semi O.~Ganor,
hep-th/9903110 \semi H.~Braden, A.~Marshakov, A.~Mironov,
A.~Morozov, hep-th/9812078}

\lref\nok{N.~Nekrasov, A.~Okounkov, to appear}

\lref\cmn{S.~Cherkis, G.~Moore, N.~Nekrasov, in progress}

\lref\dijkgraaf{R.~Dijkgraaf, hep-th/9609022}

\lref\iqbal{A.~Iqbal, hep-th/0212279}

\lref\lerche{ A.~Klemm, W.~Lerche, P.~Mayr, C.~Vafa, N.~Warner,
hep-th/9604034}

\lref\wittenm{E.~Witten, hep-th/9503124}

\lref\niklos{A.~Losev, N.~Nekrasov, in progress}

\lref\experiment{G.~Chan, E.~D'Hoker, hep-th/9906193 \semi
E.~D'Hoker, I.~Krichever, D.~Phong, hep-th/9609041\semi
J.~Edelstein, M.~Gomez-Reino, J.~Mas, hep-th/9904087 \semi
J.~Edelstein, M.~Mari\~no, J.~Mas hep-th/9805172 }

\lref\flow{I.~Klebanov, N.~Nekrasov, hep-th/9911096\semi
J.~Polchinski, hep-th/0011193}

\lref\todalit{K.~Ueno, K.~Takasaki, Adv. Studies in Pure Math.
{\bf 4} (1984) 1}

\lref\kharchev{For an excellent review see, e.g. S.~Kharchev,
hep-th/9810091}

\lref\gkmmm{
A.Gorsky, I.Krichever, A.Marshakov, A.Mironov and A.Morozov,
Phys. Lett.  {\bf B355} (1995) 466; hep-th/9505035. }

\lref\witdonaldson{E.~Witten, \cmp{117}{1988}{353}}

\lref\swi{N.~Nekrasov, hep-th/0206161} \lref\fucito{U.~Bruzzo,
F.~Fucito, J.F.~Morales, A.~Tanzini, hep-th/0211108\semi
D.Bellisai, F.Fucito, A.Tanzini, G.Travaglini, hep-th/0002110,
hep-th/0003272, hep-th/0008225} \lref\flume{R.~Flume,
R.~Poghossian, hep-th/0208176\semi R.~Flume, R.~Poghossian,
H.~Storch, hep-th/0110240, hep-th/0112211} \lref\khoze{ N.~Dorey,
T.J.~Hollowood, V.~Khoze, M.~Mattis, hep-th/0206063, and
references therein}

\lref\polyakov{A.~Polyakov, hep-th/9711002, hep-th/9809057}
\lref\ads{J.~Maldacena, hep-th/9711200\semi S.~Gubser,
I.~Klebanov, A.~Polyakov, hep-th/9802109\semi E.~Witten,
hep-th/9802150}

\lref\kachru{S. Kachru, A. Klemm, W. Lerche, P. Mayr, C. Vafa,
hep-th/9508155\semi S.~Kachru, C.~Vafa, hep-th/9505105}

\lref\mmm{A.~Losev, G.~Moore, S.~Shatashvili, hep-th/9707250\semi
N.~Seiberg, hep-th/9705221}

\lref\dbranes{J.~Polchinski, hep-th/9510017}

\chardef\tempcat=\the\catcode`\@ \catcode`\@=11
\def\cyracc{\def\u##1{\if \i##1\accent"24 i
\else \accent"24 ##1\fi }}
\newfam\cyrfam \font\tencyr=wncyr10
\def\cyr{\fam\cyrfam\tencyr\cyracc}


\Title{\vbox{\baselineskip 10pt \hbox{} \hbox{ITEP-TH-18/03}
\hbox{MPIM-2003-26}       \hbox{FIAN/TD-05/03} \hbox{IHES-P/03/09}
}  } {\vbox{\vskip -30 true pt
\smallskip
   \centerline{\bf SMALL INSTANTONS, LITTLE STRINGS}
\smallskip\smallskip\centerline{\bf AND FREE
FERMIONS} \vskip2pt}}   \centerline{Andrei S. Losev$^{1,4}$,
Andrei Marshakov$^{2,3,1,4}$, Nikita A.
Nekrasov\footnote{$^{\dagger}$}{On leave of absence from: ITEP,
Moscow, 117259, Russia}$^{4}$ }
\medskip\centerline{\it $^1$ ITEP, Moscow, 117259,
Russia}\centerline{\it $^2$ Max Planck Institute of Mathematics,
Bonn, D-53072, Germany}\centerline{\it $^3$ P.N.Lebedev Physics
Institute, Moscow, 117924, Russia}\centerline{\it $^4$ IHES,
Bures-sur-Yvette, F-91440, France}
\bigskip \noindent
We present new evidence for the conjecture that BPS correlation
functions in the ${\CN}=2$ supersymmetric gauge theories are
described by an auxiliary two dimensional conformal field theory.
We study deformations of the ${\CN} =2$ supersymmetric gauge
theory by all gauge-invariant chiral operators. We calculate the
partition function of the ${\CN}=2$ theory on ${\bR}^4$ with
appropriately twisted boundary conditions. For the $U(1)$ theory
with instantons (either noncommutative, or D-instantons, depending
on the construction)  the partition function has a representation
in terms of the theory of free fermions on a sphere, and coincides
with the tau-function of the Toda lattice hierarchy. Using this
result we prove to all orders in string loop expansion that the
effective prepotential (for $U(1)$ with all chiral couplings
included) is given by the free energy of the topological string on
${\bC\bP}^1$. Gravitational descendants play an important r\^ole
in the gauge fields/string correspondence.  The dual string is
identified with the little string bound to the fivebrane wrapped
on the two-sphere. We also discuss the theory with fundamental
matter hypermultiplets.

\bigskip\bigskip\bigskip
\Date{February 2003}

\newsec{INTRODUCTION}

The Holy Grail of the theoretical physics is the nonperturbative
theory which includes quantum gravity, sometimes called  M-theory
\wittenm. The current wisdom  says there is no fundamental coupling
constant. Whatever (string) perturbation theory is used
depends on the particular solution one expands about. The
expansion parameter is one of the geometric characteristics of the
background. It is obviously interesting to look for simplified string
and field theoretic models, which have string loop expansion, and
where the string coupling constant has a geometric interpretation.

\mezzo{String expansion in gauge theory}

Large $N$ gauge theories are the most popular, and the most
elusive models with string representation. In the gauge/string
duality \polyakov\ads\ one matches the connected correlation
functions of the gauge theory observables with the partition
function of the string theory in the bulk.  The closed string dual
has $1\over N^2$ as a string coupling constant. Advances in the
studies of the type II string compactifications on Calabi-Yau
manifolds led to another class of models, which in the low-energy
limit reduce to ${\CN}=2$ supersymmetric gauge theories, with a
novel type of string loop expansion. Namely, certain couplings
${\CF}_g$ in the low-energy effective action are given by the
genus $g$ partition function of the topologically twisted string
on Calabi-Yau. The gauge group of the ${\CN}=2$ theory does not
have to be $U(N)$ with large $N$. It is determined by the geometry
of Calabi-Yau manifold \kachru\lerche\vafaengine. The r\^ole of
effective string coupling is played by the vev of the graviphoton
field strength \gravilit, which is usually assumed to be constant
\vafaoo.

\mezzo{Generalized Scherk-Schwarz construction}

In this paper we shall explain that there exists another, natural
from the gauge theory point of view, way to flesh out these
couplings. The idea is to put the theory in a nontrivial geometric
background, which we presently describe.  Namely, consider any
Lorentz-invariant field theory in $d$ dimensions. Suppose the
theory can be obtained by Kaluza-Klein reduction from some theory
in $d+1$ dimensions. In addition, suppose the theory in $d+1$
dimensions had a global symmetry group $H$. Now compactify the
$d+1$ dimensional theory on a circle ${\bf S}^1$ of circumference
$r$, with a twist, so that in going around the circle, the
space-time ${\bf R}^d$ experiences a Lorentz rotation, by an
element ${\exp} \left(r {\Omega}\right)$, and in addition a Wilson
line in the group $H$, ${\exp} \left(r {\bA}\right)$ is turned on.
The resulting theory can be now considered in the $r \to 0$ limit,
where for finite ${\Omega}, \bA$ we find extra couplings in the
$d$-dimensional Lagrangian. This is the background we shall
extensively use. More specifically we shall be mostly
interested in the four dimensional ${\CN}=2$ theories. They all
can be viewed as dimensional reductions of ${\CN}=1$ susy gauge
theories from six or five dimensions. The global symmetry group
$H$ in six dimensions is $SU(2)$ (R-symmetry).

These considerations lead to powerful results concerning exact
non-perturbative
calculations in the supersymmetric gauge theories.
In particular, one arrives at the  technique of deriving effective
prepotentials of
the ${\CN}=2$ susy gauge theories with the gauge groups $U(N_1)
\times \ldots \times U(N_k)$  \swi\ (based on
\avatars\nikfive\issues\ihiggs\freck, see also related work
in \khoze\hollowood\fucito\flume). Previously, the effective
low-energy action and the corresponding prepotential
${\CF}^{SW}$ was determined using the constraints of holomorphy
and electro-magnetic duality \sw\seiberghol\swsol.

\mezzo{Higher Casimirs in gauge theory}

One of the goals of the present paper is to extend the method
\swi\ to get the correlation functions of ${\CN}=2$ {\it chiral}
operators. This is equivalent to solving for the effective
prepotential of the ${\CN}=2$ theory whose microscopic
prepotential (see \sw\ for introduction in ${\CN}=2$ susy) is
given by: \eqn\mcrsc{{\CF}^{UV} = {\tau}_0 {\Tr} {\Phi}^2 +
\sum_{{\vec n}} {\tau}_{\vec n} \prod_{J=1}^{\infty} {1\over
n_{J}!} \left( {1\over J} {\Tr} {\Phi}^J \right)^{n_J}} where
$\vec n = (n_1, n_2, \ldots)$ label all possible gauge-invariant
polynomials in the adjoint Higgs field ${\Phi}$ (note that
${\tau}_{0,1,0,\ldots}$ shifts ${\tau}_0$). Let ${\vec\rho} =
(1,2,3,\ldots)$,  $\vert{\vec n}\vert = \sum_J n_J$, and $\vec n
\cdot \vec \rho = \sum_J J n_J$.

In order for the theory defined by \mcrsc\
to avoid vanishing of the second derivatives of prepotential
at large (quasiclassical) values of the Higgs field
\eqn\vcavrg{\langle {\Phi} \rangle_{a} \sim a \gg {\Lambda} \sim
e^{2\pi i {\tau}_0} }
and not to run into strong coupling singularity,
the couplings ${\tau}_{\vec n}$ should be
treated formally. One could also worry about the nonrenormalizabilty of the
perturbation \mcrsc. This is actually not so, provided the conjugate
prepotential ${\bar\CF}$ is kept classical
${\bar\tau}_0 {\Tr} {\bar\Phi}^2$. The action is no longer real,
however, the effective dimensions of the fields ${\Phi}$ and
${\bar\Phi}$ become $0$ and $2$, thereby justifying an infinite
number of terms in \mcrsc.

We should  note that there are relations between the deformations
generated by derivatives w.r.t. ${\tau}_{\vec n}$, which originate
in the fact that there are polynomial relations between the
single-trace operators ${\Tr} {\Phi}^J$ for $J > N$ and the
multiple-trace operators. When instantons are included these
classical relations are modified. It seems convenient to keep all
${\tau}_{\vec n}$ as independent couplings. The classical
prepotential then obeys  additional constraints: the
$N$-independent non-linear ones: \eqn\trvc{ {\p {\CF}^{UV}\over{\p}
{\tau}_{\vec n}}  = {{\p} {\CF}^{UV}\over {\p}{\tau}_{\vec
n_1}}\ldots{{\p} {\CF}^{UV}\over{\p}{\tau}_{\vec n_k}}, \qquad {\vec n} =
{\vec n}_1 + \ldots {\vec n}_k} and the $N$-dependent linear ones:
\eqn\nntrvc{\sum_{{\vec n}: \ {\vec n} \cdot {\vec \rho} = N + k}
(-1)^{\vert {\vec n} \vert} {\p \over {\p {\tau}_{\vec n}}}
{\CF}^{UV} = 0\ , \qquad k > 0} The quantum effective prepotential
obeys instanton corrected constraints \issues, which we implicitly
determine in this paper.

\mezzo{Contact terms}

The constraints of holomorphy and electro-magnetic duality are
powerful enough to determine the effective low-energy prepotential
${\CF}^{IR}$ (see \issues), up to a diffeomorphism of the
couplings ${\tau}_{\vec n}$, i.e. up to contact terms. In order to
fix the precise mapping between the microscopic couplings (which
we also call ``times'', in accordance with the terminology adopted
in integrable systems) and the macroscopic ones, one needs more
refined methods (see \freck\ for the discussion of the contact
terms and their relation to the topology of the compactifications
of the moduli spaces). As we shall explain in this paper, the
direct instanton calculus is powerful enough to solve for
${\CF}^{IR}$: \eqn\gnfnc{{\CF}^{IR} ( a, {\tau}_{\vec n} ) =
{\CF}^{SW} (a ; {\tau}_0 ) + \sum_{\vec n} {\tau}_{\vec n}
{\CO}_{\vec n} (a) + \sum_{\vec n, \vec m } {\tau}_{\vec n}
{\tau}_{\vec m} {\CO}_{\vec n\vec m} (a) + \ldots} where
\eqn\vvs{{\CO}_{\vec n}(a) = \left< \prod_{J=1}^{\infty}{1\over
n_{J}!}  \left( {1\over J}{\Tr} {\Phi}^{J} \right)^{n_J}
\right>_{a}} while ${\CO}_{\vec n\vec m}$ are the expectation
values of the contact terms between ${\CO}_{\vec n}$ and
${\CO}_{\vec m}$ \issues\moorewitten\whitham.

\vfill\eject
\mezzo{Dual/little string theories}

We shall argue that the generalized in this way prepotential
\gnfnc, which is also a generating function of the correlators of
chiral observables, is encoded in a certain {\it stringy}
partition function. We shall demonstrate that the generating
function of the expectation values of the chiral observables in
the special ${\CN}=2$ supergravity background are given by the
exponential of the all-genus partition function of the topological
string. For the $"U(1)"$ theory the dual string lives on
${\bC\bP}^1$ (A-model). To prove this we shall use the recent
results of A.~Okounkov and R.~Pandharipande who related the
partition function of the topological string on ${\bC\bP}^1$ with
the tau-function of the Toda lattice hierarchy. The expression of
the generating function of the chiral operators through the
tau-function of an integrable system is a straightforward
generalization of the experimentally well-known relation between
the Seiberg-Witten prepotentials and quasiclassical tau-functions
\gkmmm\ (see also \booksSW\ and references therein). For the
tau-function giving the generating function for the correlators of
chiral operators we will present a natural representation in terms
of free fermionic or bosonic system. We think this is a
substantial step towards the understanding the physical origin of
the results of \gkmmm.

One may think that this result is yet another example of the local
mirror symmetry \vafaengine. We should stress here that it is by
no means obvious. Indeed, a powerful method to embed ${\CN}=2$
gauge theories into string theory is by considering type II string
on local Calabi-Yau manifolds. Almost all of the results obtained
in this way can be viewed as a degeneration of the theory which
exists for global, compact Calabi-Yau manifolds. In other words,
one assumes that the gauge theory decouples from gravity and
excited string modes, when the Calabi-Yau is about to develop some
singularity, and the global structure of Calabi-Yau is not
relevant; but in this way one cannot really discuss the higher
Casimir deformations \mcrsc. However, the main claim is there: the
prepotential of the gauge theory, as well as the higher couplings
${\CF}_g$, are given by the topological string amplitudes on the
local Calabi-Yau.

If the local Calabi-Yau can be viewed as a degeneration of the
compact Calabi-Yau then one simply takes the limit of the
corresponding topological string amplitudes (effectively all
irrelevant K\"ahler classes in the A-model are sent to infinity,
and the worldsheet instantons do not know about them; however, one
has to renormalize the zero instanton term). In this case one can,
in principle, take the mirror theory, the B-model on a dual
Calabi-Yau manifold, and try to perform the analogous degeneration
there \vafaengine, this way even leads to some equations on
${\CF}_g$'s \klemmzaslow. However, the situation here is still
unsatisfactory. For the global Calabi-Yau's the whole sum
${\sum}_g {\CF}_g {\hbar}^{2g-2}$ is identified with the logarithm
of the partition function of the effective "closed string field
theory" -- the Kodaira-Spencer theory \bcov\ on the B-side
Calabi-Yau manifold. Nothing of this sort is known for the
degenerations corresponding to local Calabi-Yau's on the A-side,
for $g > 0$. For genus zero amplitudes the pair: (mirror
Calabi-Yau manifold, a holomorphic three-form) is replaced by the
pair: (an effective curve, a meromorphic 1-form) which captures
correctly the relevant periods. In \lerche\ these curves (which
are Seiberg-Witten curves of the gauge theory) were identified in
the following way. One views the degeneration of the mirror
Calabi-Yau as an ALE fibration over a base ${\bC\bP}^1$. To this
fibration one can associate a finite cover of ${\bC\bP}^1$
associated with the monodromy group of the $H_2(ALE, {\bZ})$
bundle over ${\bC\bP}^1 \backslash $ degeneration locus. This
finite cover is the Seiberg-Witten curve. In \swi\ it was
conjectured that the Kodaira-Spencer theory should become a single
free fermion theory on this curve. In the case of Calabi-Yau being
an elliptic curve this conjecture was studied long time ago
\bcov\dijkgraaf\dougcft\ with the applications to the string
theory of two dimensional Yang-Mills \tdym\ in mind.

Our contribution to the subject is the identification of the
analogue of the Kodaira-Spencer theory, at least in the specific
context we focus on in this paper. This is, we claim, the free
fermionic (or free bosonic) theory on a Riemann surface (a sphere
for $U(1)$ gauge group), in some specific ${\bW}$-background (i.e.
with the higher spin chiral operators turned on) \walg. Note that in the
conventional approach to Kodaira-Spencer theory via type B topological
strings, the $\bW$-deformations are not considered (except for the
${\bW}_2$, corresponding to the complex structure deformations). This is a
mirror to the fact that on the A-side one does not usually takes into
account the contribution of the gravitational descendents.

Thus, we also have something new on the A-side. It is of
course not the first time when the Fano varieties appear in the
context of local mirror symmetry. However, the topological string
amplitudes, corresponding to the local Calabi-Yau do not coincide
with those for Fano, even if the actual worldsheet instantons land
on Fano subvariety in the local Calabi-Yau. For example, the
resolved conifold is the ${\CO}(-1)\oplus{\CO}(-1)$ bundle over
${\bC\bP}^1$, all worldsheet instantons land in ${\bC\bP}^1$
(which is Fano), yet the topological string amplitude is affected
by the zero modes of the fermions, corresponding to the normal
directions. These zero modes make the contributions of all
positive ghost number observables of topological string on Fano
vanish when Fano is embedded into Calabi-Yau.

In our case, however, we get literally strings on ${\bC\bP}^1$. This model
is much richer
then the strings on conifold. In particular, as we show, the gravitational
descendants of the K\"ahler class of ${\bC\bP}^1$ are dual to the higher
Casimirs in the gauge theory.

It goes without saying that embedding our picture in the general story
of local mirror symmetry
will be beneficial for both. In particular, \bcov\ explains how the
topological string amplitudes arise as the physical string amplitudes
with the insertion of $2g$ powers of the sugra Weyl multiplet ${\CW}$,
the vertex operators for ${\CW}$ effectively twisting the worldsheet
theory.
We claim that the topological string with the gravitational descendants
(which are constructed with the help of the fields of topological gravity)
have direct and clear physical meaning on the gauge theory side. We do not
know at the moment how to embed them in the framework of \bcov. However
we shall make a suggestion.

\mezzo{Organization of the paper}

The paper is organized as follows. The section $2$ discusses
instanton calculus in the ${\CN}=2$ susy gauge theories from
the physical point of view. The
mathematical aspects, related to the equivariant cohomology of the
moduli spaces and the equivariant methods which lead to the evaluation of
the integrals one encounters in the gauge theory are described in the appendix {\bf A}.
As a result of these calculations one
arrives at the generating function of the expectation values of
the chiral operators, which is expressed as a partition function
of a certain auxiliary statistical model on the Young diagrams.
The section $3$ specifies these results for the gauge group $U(1)$
and explains their interpretation from the point of view of the
little string theory, which we claim is equivalent in this case to the
topological string on ${\bP}^1$, with the gravitational descendendents of
the K\"ahler form ${\s}_k ({\o})$ lifted to the action.

This section also introduces the formalism
of free fermions which are very efficient in packaging the sums
over partitions. The section $4$ identifies the partition function
with a simple correlator of free fermions, and also with the tau-function of the
Toda lattice.   The section $5$
discusses the theory with fundamental matter, and its free field
realization.

\newsec{${\CN}=2$ THEORY}

\subsec{Gauge theory realizations}

We start our exposition with the case of pure ${\CN}=2$
supersymmetric Yang-Mills theory with the gauge group $U(N)$ and
its maximal torus ${\bf T} = U(1)^N$. The field content of the
theory is given by the vector multiplet ${\bf\Phi}$, whose
components are: the complex scalar ${\Phi}$, two gluions
${\l}_{\a}^{i}$, $i=1,2$; ${\a}=1,2$ their conjugates
${\bar\l}_{\dot \a i}$, and the gauge field $A_{\m}$ -- all fields
in the adjoint representation of $U(N)$. The action is given by
the integral over the superspace:
\eqn\ssyact{S \propto \int d^4 x \left( \int d^4 {\t} {\CF}
({\bf\Phi}) + \int d^4 {\bar\theta} {\bar\CF}({\bar{\bf\Phi}})
\right)}
where ${\t}_{\a}^i$, ${\a}=1,2$; $i=1,2$ are the chiral Grassmann
coordinates on the superspace, ${\bf\Phi} = {\Phi} + {\t}{\l} +
{\t}{\t} F^{-} + \ldots$ is the ${\CN}=2$  vector superfield, and
${\CF}$ is the prepotential (locally, a holomorphic gauge
invariant function of ${\Phi}$). Classical supersymmetric
Yang-Mills theory has
\eqn\prpcl{{\CF}({\Phi}) = {\tau}_0 {\Tr} {\Phi}^2}
where ${\tau}_0$ is a complex constant, whose real and imaginary parts
give the theta angle and the inverse square of the gauge coupling
respectively:
\eqn\brcp{{\tau}_0 = {{\vartheta}_0\over{2\pi}} + {4{\pi i}\over g_0^2} , }
the subscript $0$ reminds us that these are bare quantities, defined at
some high energy scale ${\m}_{UV}$.
 It is well-known that ${\CN}=2$ gauge theory has a
moduli space of vacua, characterized by the expectation value of
the complex scalar ${\Phi}$ in the adjoint representation. In the
vacuum $[{\Phi}, {\bar\Phi} ] =0$,   due to the potential term
${\Tr} [{\Phi}, {\bar \Phi}]^2$ in the action of the theory. Thus,
one can gauge rotate ${\Phi}$ to the Cartan subalgebra of $\bg$:
$\langle \Phi \rangle = a \in {\bf t} = Lie ({\bf T})$. We are
studying the gauge theory on Euclidean space ${\bR}^4$, and impose
the boundary condition ${\Phi} (x) \to a$, for $x \to \infty$. It
is also convenient to accompany the fixing of the asymptotics of
the Higgs field  by the fixing the allowed gauge transformations
to approach unity at infinity.

The ${\CN}=2$ gauge theory in four dimensions is a dimensional
reduction of the ${\CN}=1$ five dimensional theory. The latter
theory needs an ultraviolet completion to be well-defined.
However, some features of its low-energy behavior are robust
\seibergfive.

In particular, the effective gauge coupling runs because of the
one-loop vacuum polarization by the BPS particles. These particles
are W-bosons (for nonabelian theory),  four dimensional instantons,
viewed as solitons in five dimensional theory,  and the bound
states thereof.

To calculate the effective couplings we need to know the
multiplicities, the masses, the charges, and the spins of the BPS
particles present in the spectrum of the theory
\nikfive\gopakumarvafa. This can be done, in principle, by careful
quantization of the moduli space of collective coordinates of the
soliton solutions (which are four dimensional gauge instantons).
Now suppose the theory is compactified on a circle. Then the
one-loop effect of a given particle consists of a bulk term,
present in the five dimensional theory, and a new finite-size
effect, having to do with the loops wrapping the circle in
space-time \nikfive. If in addition the noncompact part of the
space-time in going around the circle is rotated then the loops
wrapping the circle would have to be localized near the origin in
the space-time. This localization is at the core of the method we
are employing. Its mathematical implementation is discussed in the
next section. Physically, the multiplicities of the BPS states are
accounted for by the supersymmetric character-valued index
\gopakumarvafa:
$$
\sum_{solitons} {\Tr}_{\CH} (-)^F e^{-r {\bP}_{5}}
e^{r \Omega\cdot {\bM}} e^{r \bA \cdot {\bI}}
$$
where ${\bP}_5$ is the momentum in the fifth direction, $\bM$ is the
generator of the Lorentz rotations,  $\bI$ is the generator of
the R-symmetry rotations, and $r$ is the circumference of the fifth circle.
Under certain conditions on $\Omega$ and
$\bA$ this trace has some supersymmetry which allows to evaluate it.
In the process one gets some integrals over the
instanton collective coordinates, as in
\cecotti\smilga\sethi\ihiggs. As in \ihiggs\ these integrals
are exactly calculable, thanks to the equivariant localization, described
in appendix.

Another point of view on our method is that by appropriately deforming the
theory (in a controllable way) we achieve that the path integral
has isolated saddle points, and thanks to the supersymmetry is exactly
given by the WKB approximation. The final answer is then the sum over these
critical points of the ratio of bosonic and fermionic determinants.
This sum is shown to be equal to the partition function of an auxilliary
statistical model, desribing the random growth of the Young diagrams.
We describe this model in detail in the section $2.7$.

We now conclude our discussion of
the reduction of the five dimensional theory down to four dimensions.
Actually,
we can be more general, and discuss the reduction from six dimensions.

Consider lifting the ${\CN}=2$ four dimensional theory to
${\CN} = (1,0)$ six dimensional theory, and then compactifying on
a two-torus with the twisted boundary conditions (along both $A$
and $B$ cycles), such that as we go around a non-contractible loop
${\ell} \sim n A + m B$, the space-time and the fields of the
gauge theory charged under the R-symmetry group $SU(2)_I$ are
rotated by the element $( e^{i ( n a_1  + m b_1){\s}_3 } , e^{ i
(n a_2 + m b_2 ){\s}_3 } , e^{i (n a_2 + m b_2){\s}_3} ) \in
SU(2)_L \times SU(2)_R \times SU(2)_{I}= Spin(4) \times
SU(2)_{I}$. In other words, we compactify the six dimensional
${\CN}=1$ susy gauge theory on the manifold with the topology
${\bT}^2 \times {\bR}^4$ with the metric and the R-symmetry gauge
field Wilson line: \eqn\twm{\eqalign{& ds^2 = r^2 dz d{\zb} +
(dx^{\m} + {\Omega}^{\m}_{\n}x^{\n} dz +
{\bar\Omega}^{\m}_{\n}x^{\n} d{\zb})^2, \cr & \qquad {\bA}^{a} =
({\Omega}^{\m\n} dz + {\bar\Omega}^{\m\n} d{\zb})
{\eta}_{\m\n}^{a}, \ {\m}=1,2,3,4, \ a=1,2,3 \cr}}where ${\eta}$
is the anti-self-dual 't Hooft symbol. It is convenient to combine
$a_{1,2}$ and $b_{1,2}$ into two complex parameters ${\e}_{1,2}$:
\eqn\eps{{\e}_1 - {\e}_2 = 2(a_1 + i b_1), \qquad {\e}_1 + {\e}_2
= 2(a_2 + i b_2)} The antisymmetric matrices ${\Omega},
{\bar\Omega}$ are given by: \eqn\omgs{{\Omega}^{\m\n} = \pmatrix{
0 & {\e}_1 & 0 & 0 \cr -{\e}_1 & 0 & 0 & 0 \cr 0 & 0 & 0 & {\e}_2
\cr 0 & 0 & - {\e}_2 & 0 \cr}, \qquad {\bar\Omega}^{\m\n} =
\pmatrix{ 0 & {\bar\e}_1 & 0 & 0 \cr -{\bar\e}_1 & 0 & 0 & 0 \cr 0
& 0 & 0 & {\bar\e}_2 \cr 0 & 0 & - {\bar\e}_2 & 0 \cr}} Clearly,
$[{\Omega}, {\bar\Omega}] = 0$. In the limit $r \to 0$ we get four
dimensional gauge theory. We could also take the limit to the five
dimensional theory, by considering the degenerate torus ${\bT}^2$.
We note in passing that the complex structure of the torus
${\bT}^2$ could be kept finite. The resulting four dimensional
theory (for gauge group $SU(2)$) is related to the theory of the
so-called E-strings \estring\ganor. The instanton contributions to
the correlation functions of the chiral operators in this theory
are related to the elliptic genera of the instanton moduli space
\cstw\ and could be summed up, giving rise to the Seiberg-Witten
curves for these theories. However, in this paper we shall neither
discuss elliptic, nor trigonometric limits, even though they lead
to interesting integrable systems \fivedim.

The action of the four dimensional theory in the limit $r \to 0$
is not that of the pure supersymmetric Yang-Mills theory on
${\bR}^4$. Rather, it is a deformation of the latter by the
${\Omega}$, ${\bar\Omega}$-dependent terms. We shall write down
here only the terms with bosonic fields (for simplicity, we have
set $\vartheta_0=0$):
\eqn\dfrms{S ({\Omega})^{bos} = -{1\over{2g_0^2}} {\Tr} \left(
{\half} F_{\m\n}^2 + ( D_{\m}{\Phi} - {\Omega}^{\n}_{\l} x^{\l}
F_{\m\n} ) (D_{\m}{\bar\Phi} - {\bar\Omega}^{\n}_{\l}
x^{\l}F_{\m\n} ) + [{\Phi},{\bar\Phi}]^2 \right)}
We shall call the theory \dfrms\ an ${\CN}=2$ theory in the
${\Omega}$-background.
It is amusing that this deformation can be indeed described as a
superspace-dependent bare coupling ${\tau}_0$:
\eqn\sspdpc{{\tau}_0 (x, {\theta}; {\m}_{UV}) = {\tau}_0
({\m}_{UV}) + {\bar\Omega}^{-} {\t}{\t} + {\Omega}_{\m\n}
{\bar\Omega}_{\m\l} x^{\n} x^{\l}} We are going to study the
correlation functions of chiral observables. These observables are
gauge invariant  holomorphic functions of the superfield
${\bf\Phi}$. Viewed as a function on the superspace, every such
observable ${\CO}$ can be decomposed:
\eqn\dcmps{{\CO}[{\bf\Phi}(x,{\t})] = {\CO}^{(0)} +
{\CO}^{(1)}{\t}+\ldots + {\CO}^{(4)} {\t}^4} The component
${\CO}^{(4)}$ can be used to deform the action of the theory, this
deformation is equivalent to the addition of ${\CO}$ to the bare
prepotential.

The nice property of the chiral observables is the independence of
their correlation functions of the anti-chiral deformations of the
theory, in particular of ${\bar\tau}_0$\foot{However, beware of
the holomorphic anomaly.}. We can, therefore, consider the limit
${\bar\tau}_0 \to \infty$. In this limit the term:
$$
{\bar\tau}_0 \Vert F^{+} \Vert^2
$$
in the action localizes the path integral onto the instanton
configurations. In addition, the $\Omega$-background further localizes
the measure on the instantons, invariant under rotations. Finally, the vev of the Higgs
field shrinks these instantons to the points, thus eliminating all
integrations, reducing them to the single sum over the point-like invariant
instantons.

Now we want to pause to discuss other physical realizations of our
${\CN}=2$ theories.

\subsec{String theory realizations}

The ${\CN}=2$ theory can arise as a low energy  limit of the theory on
a stack of D-branes in type II gauge theory. A stack of $N$ parallel
D3 branes in IIB theory in flat ${\bR}^{1,9}$ carries ${\CN}=4$
supersymmetric Yang-Mills theory \witbound.
A stack of parallel D4 branes in IIA theory in flat ${\bR}^{1,9}$
carries ${\CN}=2$ supersymmetric Yang-Mills theory in five dimensions.
Upon compactification on a circle the latter theory reduces to the former
in the limit of zero radius.

Now consider the stack of $N$ D4 branes in the geometry
${\bS}^1 \times {\bR}^{1,8}$
with the metric:
\eqn\stck{ds^2 = dx^{\m} dx^{\m} + r^2 d{\varphi}^2 + d v^2   +
\vert dZ_1 + m r Z_1 d{\varphi} \vert^2 +
 \vert d Z_2 - m r Z_2 d{\varphi} \vert^2}
Here $x^{\m}$ denote the coordinates on the Minkowski space
${\bR}^{1,3}$, ${\varphi}$ is the periodic coordinate on the
circle of circumference $r$, $v$ is a real transverse direction, $
Z_1$ and $Z_2$  are the holomorphic coordinates on the remaining
${\bC}^2$. The worldvolume of the branes is ${\bS}^1 \times
{\bR}^{1,3}$, which is located at $Z_1 = Z_2 = 0$, and $v=v_l$,
$l=1, \ldots, N$. Together with the Wilson loop eigenvalues
$e^{i{\s}_1}, \ldots, e^{i{\s}_N}$ around ${\bS}^1$ $v_l$'s  form
$N$ complex moduli $w_1, \ldots, w_N$, parameterizing the moduli
space of vacua. In the limit $r \to 0$ the $N$ complex moduli
loose periodicity.

It is easy to check that the worldvolume theory has ${\CN}=2$
susy, with the massive hypermultiplet in the adjoint
representation (of mass $m$). This realization is T-dual to the standard
realization with the NS5 branes \wittensolution\foot{NN thanks
M.~Douglas for the illuminating discussion on this point.}. Note
that the background \twm\ is similar to \stck. However,  the
D-branes are differently located, the fact which leads to very
interesting geometries upon T-dualities and lifts to M-theory
\cmn, providing (hopefully) another useful insight.

However, in our story we want to analyze the pure ${\CN}=2$
supersymmetric Yang-Mills theory. This can be achieved by taking
$m \to \infty$ limit, at the same time taking the weak string
coupling limit. The resulting brane configuration can be described
using two parallel NS5 branes and $N$ D4 brane suspended between
them, as in \wittensolution, or, alternatively, as a stack of $N$
D3 (fractional) branes stuck at the ${\bC}^2/{\bZ}_2$ singularity,
as in \frbranes. In fact the precise form of the singularity is irrelevant, as long as it
corresponds to a discrete subgroup of $SU(2)$, and all the fractional branes are of the same
type. Note that for $m r = {1\over
K}$ the
$(Z^1, Z^2)$ part of  the metric \stck\ in the limit $r \to 0$
looks like the metric on the orbifold ${\bC}^2/{\bZ}_K$.
  The relation between these two pictures is
through the T-duality of the resolved ${\bC}^2/{\bZ}_2$
singularity. The fractional D3 branes blow up into D5 branes
wrapping a non-contractible two-sphere. The resolved space
$T^*{\bC\bP}^1$ has a $U(1)$ isometry, with two fixed points (the
North and South poles of the non-contractible two-sphere). Upon
T-duality these turn into two NS5 branes. The D5 branes dualize to
D4 branes suspended between NS5's.

The instanton effects in this theory are due to the fractional D(-1)
instantons, which bind to the fractional D3 branes, in the IIB description.
The ``worldvolume'' theory on these D(-1) instantons is the supersymmetric
matrix integral, which we describe with the help of ADHM construction below.
In the IIA picture the instanton effects are due to Euclidean D0 branes,
which ``propagate'' between two NS5 branes.

The IIB picture with the fractional branes corresponds to the metric
(before ${\Omega}$ is turned on):
\eqn\stckii{ds^2 = dx^{\m} dx^{\m} + dwd{\wb}   + ds^2_{{\bC}^2/{\bZ}_2}}

The singularity ${\bC}^2/{\bZ}_2$ has five moduli in IIB string
theory: three parameters of the geometric resolution of the
singularity, and the fluxes of the NSNS and RR 2-forms through the
two-cycle which appears after blowup. The latter are responsible
for the gauge couplings on the fractional D3 branes \cftorb:
\eqn\ggcpl{{\tau}_0 = \int_{{\bS}^2} B_{RR} + {\tau}_{IIB}
\int_{{\bS}^2}B_{NSNS}}

Our conjecture is that turning on the higher Casimirs, (and
gravitational descendants on the dual closed string side)
corresponds to a ``holomorphic wave'', where ${\tau}_0$
holomorphically depend on $w$. This is known to be a solution of
IIB sugra \flow.

We shall return to the fractional brane picture later on. Right
now let us mention another stringy effect. By turning on the
constant NSNS B-field along the worldvolume of the D3-branes we
deform the super-Yang-Mills on ${\bR}^4$ to the super-Yang-Mills
on the noncommutative ${\bR}^4_{\Theta}$ \cds\witsei\connes. On
the worldvolume of the D(-1) instantons the noncommutativity acts
as a Fayet-Illiopoulos term, deforming the ADHM equations
\abs\abkss\neksch, and resolving the singularities of the
instanton moduli space, as in \nakajima. We shall use this
deformation as a technical tool, so we shall not describe it in
much detail. The necessary references can be found in \witsei.

At this point we remark that even for $N=1$ the instantons are
present in the D-brane picture. They become visible in the gauge
theory when noncommutativity is turned on. Remarkably, the actual
value of the noncommutativity parameter ${\Theta}$ does not affect
the expectation values of the chiral observables, thus simplifying
our life enormously.

So far we presented the D-brane realization of ${\CN}=2$ theory. There
exists another useful realization, via local Calabi-Yau manifolds
\vafaengine. This realization, as we already explained
in the introduction is useful in relating the prepotential to the
topological string amplitudes.
If the theory is embedded in the IIA string on local Calabi-Yau,
then the interesting
physics comes from the worldsheet instantons, wrapping some 2-cycles
in the Calabi-Yau.
In the mirror IIB description one gets a string without worldsheet
instantons contributing
to the prepotential, and effectively reducing to some field theory.
This field theory is known in the case of global Calabi-Yau. But it
is not known explicitly in the case of local Calabi-Yau.
As we shall show, it can be sometimes identified with the free fermion
theory on auxiliary Riemann surface (cf. \dijkgraaf).

Relation to the geometrical engineering \vafaengine\ is also useful in
making contact between our ${\Omega}$-deformation and the sugra backgrounds
with graviphoton field strength.
Indeed, our construction involved a lift to five or six dimensions. The
first case embeds easily to IIA string theory where
 this corresponds to the lift to M-theory. To see the whole six dimensional
picture \twm\ one should use IIB language and the lift to F-theory (one
has to set ${\bar\Omega}=0$, though).

Let us consider the five dimensional lift.
We have M-theory on the 11-fold with the metric:
\eqn\mmetrc{ds^2 = ( dx^{\m} + {\Omega}^{\m}_{\n}x^{\n} d{\varphi})^2 +
r^2 d{\varphi}^2 +  ds_{CY}^2}
Here we assume, for simplicity, that ${\e}_1 = - {\e}_2$, so that
${\Omega} = {\Omega}^{-}$ generates an $SU(2)$ rotation, thus preserving
half of susy.
Now let us reduce on the circle ${\bS}^1$ and interpret the background
\mmetrc\ in the type IIA string. Using \wittenm\ we arrive at the
following IIA background:
\eqn\iias{\eqalign{& g_s = \left( r^2 + \Vert \Omega \cdot x \Vert^2
\right)^{3\over 4} \cr
& A^{grav} = {1\over{r^2 + \Vert \Omega \cdot x \Vert^2}} {\Omega}_{\m\n}
x^{\m} dx^{\n} \cr
& ds_{10}^{2} = {1\over{\sqrt{r^2 + \Vert \Omega \cdot x \Vert^2}}}
\left( r^2 dx^2 + {\Omega}^{\m}_{\n}{\Omega}^{\l}_{\k} \left( x^2 dx^2
{\d}^{\n\k}{\d}_{\m\l} - x^{\n}x^{\k} dx^{\m} dx^{\l} \right)\right) +
\cr & \qquad \qquad \qquad\qquad \qquad \qquad + \sqrt{ r^2 + \Vert
\Omega \cdot x \Vert^2} ds_{CY}^2 \cr}}
where the graviphoton $U(1)$ field is turned on. The IIA string coupling
becomes strong at $x \to \infty$. However, the effective coupling in the
calculations of ${\CF}_g$ is
\eqn\effcpl{{\hbar} \sim g_s \sqrt{ \Vert dA^{grav} \Vert^2} \sim
\left( r^2 +\Vert \Omega \cdot x \Vert^2\right)^{-{1\over 4}} \to 0,
\qquad x \to \infty}

\subsec{The partition function}

Our next goal is the calculation of the partition function
\eqn\partnf{Z({\tau}_{\vec n}; a, {\Omega}) =
\int_{{\phi}({\infty}) = a} D{\Phi}DA D{\l} \ldots
e^{-S({\Omega})}} of the ${\CN}=2$ susy gauge theory with all the
higher couplings \mcrsc\ on the background \twm\ with the fixed
asymptotics of the Higgs field at infinity. We use the fact that
the chiral deformations are not sensitive to the anti-chiral
parameters (up to holomorphic anomaly \niklos). We take the limit
${\bar\tau}_0 \to \infty$, and the partition function becomes the
sum over the instanton charges of the integrals over the moduli
spaces ${\CM}$ of instantons of the measure, obtained by the
developing the path integral perturbation expansion around
instanton solutions.

On the other hand, if we take instead a low-energy limit, this
calculation should reduce to that of low-energy effective theory.
In the Seiberg-Witten story \sw\ the low-energy theory is
characterized by the complexified energy scale
${\Lambda} \sim {\m}_{UV} e^{2\pi i {\tau}_0 ({\m}_{UV})}$.
We now recall \sspdpc. In our setup the low-energy scale is
$(x,{\t})$-dependent:
\eqn\irsc{{
\Lambda}(x,{\t} ) = {\m}_{UV} e^{2\pi i {\tau}_0 (x, {\t};
{\m}_{UV})} = {\Lambda} e^{2\pi i {\bar\Omega}^{-} {\t}^2 -
\Vert {\Omega} \cdot x \Vert^2}}
Near $x=0$ it is finite, while at $x \to \infty$ the theory
becomes infinitely weakly coupled. With \sspdpc\ in mind we can
easily relate the partition function to the prepotential \gnfnc
(cf. \swi): \eqn\rltns{\eqalign{& Z = Z^{pert} \ {\exp} \left[
\int d^4 x d^4 {\t} {\CF}^{inst} \left(a  ; {\tau}_{\vec n};
{\Lambda}(x,{\t}) \right)+ {\rm higher \ derivatives} \right] =
\cr & \qquad = {\exp}\ {1\over{{\e}_1 {\e}_2}} \left[  {\CF} (a,
{\tau}_{\vec n}; {\Lambda})+ O ({\e}_1, {\e}_2) \right]\cr}} where
${\CF}^{inst}$ is the sum of all instanton corrections to the
prepotential, and $Z^{pert}$ is the result of the perturbative
calculation on the ${\Omega}$-background. The corrections in
${\e}_{1,2}$ come from the ignored higher derivative terms.

\subsec{Perturbative part}

The
perturbative part is given by the one-loop contribution from
W-bosons, as well as non-zero angular momentum modes of the abelian photons
(we shall comment on this below).
Recall
that in the $\Omega$-background one can integrate out all non-zero
modes, as ${\Omega}$ lifts all massless fields. Because of the
reduced supersymmetry the determinants do not quite cancel. The
simplest way to calculate them is to go to the basis of
normalizable spherical harmonics: \eqn\sphrc{ {\Phi} = \sum_{l,m
=1}^{N} T_{lm} \ \sum_{i,j, {\ib}, {\jb} \geq 1}
{\phi}^{lm}_{ij{\ib}{\jb}}\
 z_1^{i-1}z_2^{j-1} {\zb}_1^{{\ib}-1}{\zb}_2^{{\jb}-1} \ e^{-\vert z_1 \vert^2 - \vert z_2
 \vert^2}}and similarly for the components of the gauge fields and
 so on. Here the terms with $l \neq m$ correspond to the W-bosons,
massive components of the Higgs field, and the massive components of the
gluinos, while $l=m$ represent the abelian part.
We are doing the WKB calculation around the trivial gauge field
$A=0$: the unborken susy guarantees there are no further corrections.
The integral over the bosonic and fermionic fluctuations becomes a
ratio of the determinants, formally:
\eqn\wghts{\prod_{l,m=1}^{N} \prod_{i,j =
1}^{\infty} \prod_{{\ib}, {\jb}=1}^{\infty} {{(a_{lm} + {\e}_1 ( i
- {\ib}) + {\e}_2 ( j - {\jb} ))(a_{lm} + {\e}_1  ( i - {\ib} - 1)
+ {\e}_2 ( j - {\jb} - 1))}\over{(a_{lm} + {\e}_1 ( i - {\ib}-1) +
{\e}_2 ( j - {\jb} ))(a_{lm} + {\e}_1  ( i - {\ib}) + {\e}_2 ( j -
{\jb} - 1))}}}(recall that the ``weight'' of $A_{\m}(z,{\zb})$
has in addition to its ``orbital'' weight, which comes from the $(z,
{\zb})$-dependence, a spin  $(-{\e}_1)$ weight, similarly for
$A_{\bar 2}$ we have an extra $(-{\e}_2)$, for $F^{0,2}$ extra
$(-{\e}_1 - {\e}_2)$).
In the product over ${\ib}, {\jb}$
only the term with ${\ib}={\jb}=1$
is not cancelled,
giving rise to: \eqn\prt{Z^{pert} =\ {\prod_{l,m; i,j \geq
1}}^{\kern -.12in\prime}\ \left( a_l - a_m + {\e}_1 (i-1) + {\e}_2
(j-1) \right) } times the conjugate term, which depends on $\bar
a$. We shall ultimately take $\bar a \to\infty$, so we ignore this
term -- at any rate, it cancels out in the correlation functions of
the chiral observables. The symbol ${\prod}^\prime$ in \prt\
means that the contribution of the abelian zero angular momentum modes $l=m$, $i=j=1$ to
the product is omitted (this has to do with our boundary
conditions). We shall always understand \prt\ in the sense of
$\zeta$-regularization. After regularization one can analytically
continue to ${\e}_1 + {\e}_2= 0$.

In fact, for ${\e}_1 = - {\e}_2
= {\hbar}$ one can expand:
\eqn\gns{Z({\tau}_{\vec n}; a, {\Omega}) = {\exp} \left(-
\sum_{g=0}^{\infty} {\hbar}^{2g-2} {\CF}_{g} (a; {\tau}_{\vec n};
{\Lambda})\right)} The higher ``prepotentials'' ${\CF}_{g}$ will
turn out later to be related to the higher genus string
amplitudes.

\subsec{Mathematical realization of ${\CN}=2$ theory}

The mathematical realization of the gauge theory we are studying
is the following (details are in the appendix A). Consider the
space ${\CY}$ of all gauge fields on ${\bR}^4$ with finite
Yang-Mills action. There are three groups of symmetries acting on
this space which we shall study. The first group,
${\CG}_{\infty}$, is the group of gauge transformations, trivial
at infinity: $g(x) \to 1$, $x \to \infty$. The second, $G$ is the
group of constant, global, gauge transformations. The group of all
gauge transformations ${\CG}$ is the extension of ${\CG}_{\infty}$
by $G$, s.t. $G = {\CG}/{\CG}_{\infty}$. The third group $K =
Spin(4)$, is the covering group of the group of Euclidean
rotations about some fixed point $x=0$. Over the space ${\CY}$ we
consider the ${\CG} \sdtimes K$-equivariant vector bundle ${\CV}$
of the self-dual two-forms on ${\bR}^4$ with the values in the
adjoint representation of the gauge group ${\CG}$. For a gauge
field $A \in {\CY}$ the self-dual projection of its curvature
$F^{+}_{A}$ defines a section of ${\CV}$.

The path integral measure of the supersymmetric gauge theory with
the extra ${\Omega}$-couplings is nothing but the Mathai-Quillen
representative of the Euler class of ${\CV}$, written using the
section $F^+$, and working with ${\CG} \sdtimes K$ equivariantly.
Calculating the path integral corresponds to the pushforward onto
the quotient by the group ${\CG}_{\infty}$ and its further
localization w.r.t the remaining groups $G \times K$.  The result
is given by the sum over the fixed points of the $G \times K$
action on the moduli space of instantons ${\CM}$, i.e. solutions
to $F^+ = 0$.

The chiral observables translate to the equivariant Chern classes
of some natural bundles (sheaves) over the moduli space ${\CM}$.
Their calculation is more or less standard and is presented in the next section.

\subsec{Nonperturbative part}

We now proceed with the calculation of the nonperturbative contribution to
the partition function \partnf. There are two ways of determining it.
One way is the direct analysis of the saddle points of the path integral
measure. This is a nice excersize, but it relies on very explicit knowledge
of the deformed instanton solutions \neksch\cg\opennc, invariant under the
action of the group $K$ of rotations \branek. Instead, we shall choose
slightly less explicit, but more general route.

The general property of the chiral observables in ${\CN}=2$ theories,
which
is a direct consequence of the analysis in \witdonaldson, is the
cohomological nature of their correlation functions. Namely, in the limit
${\bar\tau}_0 \to 0$ these become the integrals over the instanton moduli
space $\CM$. The chiral observables, evaluated on the instanton collective
coordinates, become closed differential forms. Thus, if the moduli space
$\CM$ was compact and smooth, one could choose some convenient
representatives of their cohomology classes to evaluate their integrals.
Moreover, a generalization of the arguments in \witdonaldson\ allows to
consider the ${\CN}=2$ theory in the ${\Omega}$-background. In this case the
differential forms on $\CM$ become $K$-equivariantly closed. Even though the
space $\CM$ is not compact, the space of $K$-fixed points is, and this is
good enough for the evaluation of the integrals of the $K$-equivariant
integrals.

The final bit of information which makes the calculation of the chiral
observables constructed out of the higher Casimirs possible, is the
identification of the $K$-equivariantly closed differential forms on $\CM$
they represent with the densities of the equivariant Chern classes of some
natural bundles over $\CM$. We now proceed with the explicit description of
$\CM$, these natural bundles, and finally the chiral observables.

\vfill\eject
\mezzo{ADHM construction}

To get a handle on these fixed point
sets and to calculate the characteristic numbers of the various
bundles we have defined above, we need to remind a few facts about
the actual construction of $\CM$, the so-called ADHM construction
\adhm\nakajima. In this construction one starts with two Hermitian
vector spaces $W$ and $V$. One then looks for
four Hermitian operators ${\bX}^{\m} : V \to V$, ${\m}=1,2,3,4$
and two complex operators ${\l}_{\a} : W \to V$, ${\a}=1,2$ (and
${\bar\l}_{\dot \a} = {\l}_{\a}^{\dagger}: V \to W$), which can be
combined into a  sequence: \eqn\adhms{0 \to W \otimes S_{-}
\longrightarrow V \oplus W \otimes S_{+} \to 0} where the
non-trivial map is given by: $$ {\CD}^{+} = {\l} \oplus {\bX}^{\m}
{\s}_{\m} $$ The ADHM equation requires that ${\CD}{\CD}^{+}$
commutes with the Pauli matrices ${\s}_{\m}$ acting in $S_{-}$. In
addition, one requires that ${\CD}{\CD}^{+}$ has a maximal rank.
The moduli space ${\CM}$ is then identified with the space of such
${\bX}, {\l}$ up to the action of the group $U(k)$ of unitary
transformations in $V$. The group $G = U(N)$ acts on ${\CM}$ by the
natural action, descending from that on  ${\l}$ (${\bX}$ are
neutral). The group $K \approx Spin(4)$ acts on ${\CM}$ by rotating ${\bX}$
in the vector representation and ${\l}$ in the appropriate chiral
spinor representation.

\mezzo{D-brane picture, again}

The ADHM construction becomes very natural when the gauge theory
is realized with the help of D-branes. The space $V$ is the
Chan-Paton space for the D(-1) branes, while $W$ is the Chan-Paton
space for the D3 branes. The matrices ${\bX}$ are the ground
states of the $(-1,-1)$ strings, while ${\l}_{\a}, {\bar\l}_{\dot
a}$ are those of $(-1,3)$, $(3,-1)$. The ADHM equations are the
conditions for unbroken susy. Their solutions describe the Higgs
branch of the D(-1) instanton theory\foot{To make this statements
literally true one should consider D2-D6 system instead of
D(-1)-D3 (to avoid off-shell string amplitudes, and the
non-existence of moduli spaces of vacua in the field theories less
then in three dimensions).}. The D(-1) instantons also carry a
multiplet responsible for the $U(V)$ ``gauge'' group. In
particular, quantization of $(-1,-1)$ strings in addition to $\bX$
gives rise to a matrix $\phi$ (not to be confused with ${\Phi}$ in
the adjoint of $U(N)$!) in the adjoint of $U(k)$, which represents
the motion of D(-1) instantons in the directions, transverse to D3
branes.

\vfill\eject
\mezzo{Tangent and universal bundles.} Here we recall some standard constructions. The problem considered
here is typical in the soliton physics. One finds some
moduli space of solutions (collective coordinates) which should be
quantized. The supersymmetric theories lead to supersymmetric
quantum mechanics on the moduli spaces. If the gauge symmetry is
present the collective coordinates are defined with the help of
some gauge fixing procedure, which leads to the complications
described below.

The tangent space
to the instanton moduli space ${\CM}$ at the point $m$ can be
described as follows. Pick a gauge field $A$ which corresponds to
$m\in\CM$, $F^{+}(A)=0$. Any two such choices differ by a gauge
transformation. Now consider deforming $A$: $$A \to A + {\d} A$$
so that the new gauge field also obeys the instanton equation
$F^{+}(A + {\d} A) = 0$. In other words, ${\d}A$ obeys the linear
equations: \eqn\tngntb{\eqalign{& D^{+}_{A} {\d}A = 0 \cr &
D^{*}_{A} {\d} A = 0 \cr}} where the first equation is the
linearized anti-self-duality equation, while the second is the
gauge choice, to project out the trivial deformations ${\d}A \sim
D_{A} {\ve}$. Let us choose some basis in the (finite-dimensional)
vector space of solutions to \tngntb: ${\d} A = a_{\m}^{K} dx^{\m}
{\z}_{K}$, where $a^K$ obey \tngntb, and, say, are orthonormal
with respect to the natural metric $\langle a^L \vert a^K \rangle \equiv
\int_{{\bf R}^4} a^L \wedge \star a^K = {\d}_{LK}$, $L,K = 1,
\ldots, {\rm dim}\ {\CM}$. Now suppose we have a family of instanton
gauge fields, parameterized by the points of ${\CM}$: $A_{\m} (x;
m)$, where $x \in {\bf R}^4, \ m \in {\CM}$. Let us differentiate
$A_{\m}$ w.r.t the moduli $m$. Clearly, one can expand:
\eqn\expns{{{{\p} A}\over{{\p} m^L}} = a^{K} {\z}_{LK} + D_{A}
{\ve}_{L}} The compensating gauge transformations ${\ve}_L$
together with $A_{\m} (m)$ form a connection ${\bf\CA} = A_{\m}
(x; m) dx^{\m} + {\ve}_{L} d m^L$ in the rank $N$ vector bundle
${\CE}$ over ${\CM} \times {\bR}^4$. Now let us calculate its full
curvature:
\eqn\crtvture{{\CF} = {\bf d} {\CA} + [{\CA}, {\CA}], \qquad {\bf d} = d_{\CM} +
d_{{\bR}^4}}
\eqn\crvtexp{{\CF} = {\Phi} + {\Psi} + F}
where ${\Phi}$ is a two-form on ${\CM}$, ${\Psi}$ is a one-form on $\CM$ and
one-form on ${\bR}^4$, and $F$ is a two-form on ${\bR}^4$. The
straightforward calculation shows that ${\Phi}, {\Psi}, F$ solve the equation:
\eqn\eqhu{{\Delta}_{A} {\Phi} = [ {\Psi}, \star {\Psi}], \qquad
D^{+}_{A}{\Psi} = 0, \quad D^{*}_{A} {\Psi} = 0, \qquad F^{+}=0}The
equation on ${\Phi}$ is (up to $Q$-exact terms) identical to the equation on
the adjoint Higgs field in the instanton background, while the equation on
$\Psi$ is (again, up to $Q$-exact terms) identical to that on gluion zero
modes. This relation between ${\CF}$ and the chiral observables (which are,
after all, the polynomials in ${\Phi}, {\Psi}, F$, up to $Q$-exact terms)
will prove extremely useful in what follows. In particular, we can write:
\eqn\dscnd{\eqalign{& {\CO}_{J}^{(0)} = {1\over{J}} {\Tr}{\Phi}^J,\ \ \ \
\ldots\ ,\cr
&
{\CO}_J^{(4)} = \sum_{l=0}^{J-2} {\Tr} \left( {\Phi}^{l} F {\Phi}^{J-2-l} F
\right) + \cr
& \qquad + \sum_{l,n \geq 0, l+n \leq J-3} {\Tr} \left( {\Phi}^l F {\Phi}^n
{\Psi} {\Phi}^{J-3-l-n} {\Psi} \right) + \cr
& \qquad +\sum_{l,k,n \geq 0, l+k+n \leq J-4}
{\Tr}\left(
{\Phi}^l {\Psi}{\Phi}^k {\Psi}{\Phi}^n {\Psi}{\Phi}^{J-4-k-l-n} {\Psi}
\right)\cr}}where we substitute the expressions for $\Phi, \Psi, F$ from
\crvtexp.

A mathematically oriented
reader would object at this point, as it {\it is well-known} that
universal bundles together with a nice connections do not exist
over the compactified moduli spaces. We shall not pay attention to
these (fully just) remarks, as eventually there is a way around.
We find it more straightforward to explain things as if such
objects existed over the compactified moduli space of instantons.
Let $p$ denote the projection ${\CM} \times {\bf R}^4 \to {\CM}$.
Suppose we know everything about ${\CE}$. How would we reconstruct
$T{\CM}$ from there? We know already that the tangent space to
${\CM}$ at a point $m$ is spanned by the solutions to \tngntb. It
is plain to identify these solutions with the cohomology of the
Atiyah-Singer complex: \eqn\atsc{0 \longrightarrow
{\Omega}^{0}({\bf R}^4) \otimes {\bg} \longrightarrow
{\Omega}^{1}({\bf R}^4) {\otimes} {\bg} \longrightarrow
{\Omega}^{2,+} ({\bf R}^4) \otimes {\bg} \longrightarrow 0} where
the first non-trivial arrow is the infinitesimal gauge
transformation: ${\ve} \mapsto D_{A} {\ve}$ and the second it
${\d} A \mapsto D^{+}_{A} {\d} A$. Thanks to $F^{+}_{A} = 0$ this
is indeed a complex, i.e. $D^{+}_{A} D_{A} = 0$. The spaces
${\Omega}^{k} \otimes {\bg}$ can be viewed as the bundles over
${\CM} \times {\bf R}^4$, e.g. for $G = U(N)$
\eqn\forms{{\Omega}^{k} ({\bf R}^4) \otimes {\bg} = {\CE} \otimes
{\CE}^{*} \otimes {\Lambda}^k T^*{\bf R}^4} Generically the
complex \atsc\ has only $H^1$ cohomology. We are thus led to
identify K-classes: $T{\CM} = H^{1} - H^{0} - H^{2}$.

\mezzo{Framing  and  Dirac \ bundles.} We shall need two
more natural bundles over ${\CM}$. As ${\CM}$ is defined by the
quotient w.r.t. the group of gauge transformations, trivial at
infinity, we have a bundle $W$ over ${\CM}$ whose fiber is the
fiber of the original $U(N)$ bundle over ${\bf R}^4$ at infinity.
Another important bundle is the bundle $V$ of Dirac zero modes.
Its fiber over the point $m \in {\CM}$ is the space of
normalizable solutions to the Dirac equation in fundamental
representation in the background of the instanton gauge field,
corresponding to $m$. In  $K({\CM})$,
\eqn\kthrl{\eqalign{& W = \lim_{x \to \infty} {\CE} \vert_{x} \cr
& \qquad V = p_{*} {\CE} \cr}}The pushforward $p_{*}$ is defined here in
$L^2$ sense. In what follows we shall need its equivariant
analogue. Finally, let $S_{\pm}$ denote the bundles of positive
and negative chirality spinors over ${\bR}^4$. These bundles are
trivial topologically. However they are nontrivial as $K$-equivariant bundles.

\mezzo{Relations  among bundles.} We arrive at the
following relation among the virtual bundles: \eqn\tang{\eqalign{&
{\CE} = W \oplus V \otimes \left( S_{+} - S_{-} \right) \cr &
T{\CM} = - p_{*} \left( {\CE} \otimes{\CE}^{*} \right) \cr}} The
chiral operators ${\CO}_{\vec n}$ we discussed in the introduction
now are in one-to-one correspondence with the characteristic
classes of the $U(N)$ bundles. A convenient basis in the space of
such classes is given by the skew Schur functions, labelled by the
partitions ${\l} = ( {\l}_1 \geq {\l}_2 \geq \ldots {\l}_N \geq 0
)$: \eqn\shcf{{\bf ch}_{\l} = {\Det} \Vert ch_{{\l}_{i} - i + j}
\Vert}  Another basis is labelled by finite sequences $n_1, n_2,
\ldots, n_k$ of non-negative integers: \eqn\bsbs{{\CO}_{\vec n} =
\prod_{J=1}^{\infty} {1\over {n_{J}!}} \left({ch_J \over
J}\right)^{n_J}} It is this basis that we used in \mcrsc.

The relations \tang\ imply the relations among the Chern classes. It is
convenient to discuss the Chern characters first.
Recall that we always work $G \times K$-equivariantly.

We get:
\eqn\tangd{\eqalign{& Ch ({\CE}) = Ch(W) + Ch (V) \prod_{i=1}^2 \left( e^{x_i \over 2} - e^{-{x_i \over 2}} \right) \cr
& Ch(T{\CM}) = - \int_{{\bR}^4} Ch({\CE})Ch({\CE}^{*}) \prod_{i=1}^2 \left( {x_i \over{e^{x_i \over 2}- e^{-{x_i \over 2}}}}\right) \cr}}
where $x_1, x_2$ are the equivariant Chern roots of the tangent bundle to
${\bR}^4$:
\eqn\chrt{
x_i = {\e}_i + \CR_i}
where $\CR_i = {1\over{2\pi i}} {\d}^{2}(z_i) dz_i \wedge d{\zb}_i$
 is a curvature two-form\foot{For those worried by
the singular form of \chrt, here is a nonsingular representative. Choose a smooth function $f(r)$ which is
equal to $1$ for sufficiently large $r$, and vanishes at $r=0$.
Then $x_i$ is $K$-equivariantly cohomologous to
${\e}_i f(\vert z_i \vert^2) + {1\over{2\pi}}
f^{\prime}(\vert z_i \vert^2) dz_i \wedge d{\zb}_i$.} on ${\bR}^4$.
As everything is
$K$-equivariant, the integral over ${\bR}^4$ localizes onto the $K$-fixed
point, the origin (one also sees this from the explicit formula \chrt):
\eqn\tangch{Ch(T{\CM}) = - \left[ Ch({\CE})Ch({\CE}^*)\right]_{0} \prod_{i=1}^2 \left( {1\over{e^{{\e}_i \over 2}- e^{-{{\e}_i \over
2}}}}\right)}
where $\left[ Ch({\CE})Ch({\CE}^*)\right]_{0}$ is the evaluation of the
product of the Chern characters at the origin of ${\bR}^4$.

\mezzo{Integration over $\CM$}

Now we want to integrate over $\CM$. Suppose the integrand is the
${\bG} = G\times K$-equivariant differential form (see appendix A
for definitions) ${\Omega}_{\CO}[{\vf}]$, ${\vf} \in Lie({\bG})$.
Such integrals
can be computed using localization. In plain words it means that
there are given by the sums over the fixed points of the action of
the one-parametric subgroup ${\exp} (t {\ba}) $, $t \in {\bR}$, of
${\bG}$, ${\ba} \in Lie({\bG})$. The contribution of each fixed
point $P \in {\CM}$
(assuming it is isolated and $\CM$ is smooth at this point) is
given by the ratio: \eqn\contr{Z_{P} =
{{{\Omega}_{\CO}[{\ba}]^{(0)}\vert_{P}}
\over{c(T{\CM})[{\ba}]^{(0)}\vert_{P}}} } where ${\o}^{(0)}$
denotes the scalar component of the inhomogeneous differential
form corresponding to the equivariant differential form ${\o}$,
and $c(T{\CM})$ is the equivariant Chern polynomial of $T{\CM}$.
It is defined as follows. As $T{\CM}$ is $\bG$-equivariant, with
respect to the maximal torus $\bT$ it splits as a direct sum of
the line bundles,
$T{\CM} = \bigoplus_{i} L_i$,
on which $\bt$ acts with some weight $w_i$ (a linear function on
$\bt$). The equivariant Chern polynomial is defined simply by:
\eqn\chrpl{
c(T{\CM } )[{\ba}] = \prod_{i} \left( c_1 (L_i) + w_i \left( {\ba}
\right) \right)}
\mezzo{Physicists are familiar with the
Duistermaat-Heckmann \DHf\ formulae like \contr\ in the context of
two-dimensional Yang-Mills theory \witdgt, and in (perhaps less
known) the context of sigma models \maxim.} In order to proceed we
need to calculate the numerator and the denominator of \contr\ and
to sum over the points $P$.
We need first the equivariant Chern polynomial $c(T{\CM})$.
We already have an expression
\tangch\
for the equivariant Chern character of $T{\CM}$. To use it we recall that in
terms of $L_i$'s:
\eqn\tangchh{Ch (T{\CM}) = \sum_i e^{c_1 (L_i) + w_i \left( {\ba}
\right) }}
so that if we know \tangchh\ we also know \chrpl. Moreover, if the fixed
points $P$ are isolated (and they will be), the actual first Chern classes
of $L_i$ will never contribute (they are two-forms and we simply want to
evaluate \tangchh, \chrpl\ at a point $P$), so we only need to find $w_i$'s --
the weights.

Now, what about ${\Omega}_{\CO}$? Well, we construct it using the
descendents of the Casimirs ${\Tr}{\Phi}^J$ and their multi-trace products.
As we explained above, these become the polynomials in the traces of the
powers of the universal curvature ${\CF}$ as in \dscnd. That is to say, they
are cohomologous to the Chern classes of the universal bundle ${\CE}$.

We are mostly interested in the correlators of the $4$-descendents
${\CO}^{(4)}$ of the invariant polynomials ${\CP}({\Phi})$ on $Lie(G)$. On the moduli space
${\CM}$ these are cohomologous to the integrals over ${\bR}^4$ of the
polynomials in the Chern classes $ch_k({\CE})$ of the universal bundle.
Again, thanks to ${\bG}$-equivariance, these integrals are simply given by
the localization at the origin in ${\bR}^4$:
\eqn\forobs{{\CO}^{(4)}_{\CP} = {{\left[ {\CP} \left( {\CF} \right)
\right]_{0}}\over{{\e}_1{\e}_2}}}
For
${\CP}_k({\Phi}) ={1\over{(2\pi i)^k \ k!}}{\Tr}{\Phi}^k$,
${\CP}_k({\CF}) = ch_k({\CE})$. Any other invariant
polynomial is a polynomial in these ${\CP}_k$.

\vfill\eject
\mezzo{Evaluation of Chern classes at fixed points}

So, we see that everything reduces to the enumeration of the fixed points
$P$, and the evaluation of the Chern classes of ${\CE}$ at these points.
Moreover, thanks to \tangchh\ it is sufficient to evaluate the restriction of
$Ch(W)$ and $Ch(V)$.

These problems were solved in \swi\ for any $N$ using the results of \nakajima\ for
$N=1$.
The result is the following. The fixed points are in one-to-one
correspondence with the $N$-tuples of partitions:
${\vec\bk}  = ({\bk}_1, \ldots, {\bk}_N)$, where
\eqn\prtns{{\bk}_{l} = \left( k_{l1} \geq k_{l2} \geq k_{l3} \geq \ldots
k_{l\ n_l} > k_{l \ n_{l} +1 } = 0 \ldots \right)}
At the fixed point $P_{\vec\bk}$ corresponding to such an $N$-tuple, the
Chern characters of the bundles $W$ and $V$ evaluate to:
\eqn\chrneva{\eqalign{& \left[ Ch(W) \right]_{P_{\vec\bk}} = \sum_{l=1}^{N} e^{a_l} \cr
& \left[ Ch(V) \right]_{P_{\vec\bk}} = \sum_{l=1}^{N} \sum_{i=1}^{n_{l}} \sum_{j=1}^{k_{li}} e^{a_l +
{\e}_1 ( i-1) + {\e}_2 (j-1)}\cr}}
From this we derive an expression for $Ch({\CE})$, and for $c(T{\CM})$.

\mezzo{D-brane picture of partitions}

It is useful to recall here the D-brane interpretation of the
partitions ${\bk}$. In this picture, the fractional D3-branes are
separated in the $w$ direction, and are located at $w = a_l$,
$l=1, \ldots, N$. To the $l$'th D3 brane $k_l$
D(-1) instantons ($k_l = \sum_i k_{li}$) are attached.
In the noncommutative theory with
the noncommutativity parameter ${\Theta}$,
$$
[x^1, x^2] = [x^3, x^4] = i {\Theta}
$$
these D(-1) instantons are located near the origin $(z_1, z_2) \sim 0$,
where $z_1 = x^1 + ix^2, z_2 = x^3 + i x^4$. Different partitions
correspond to the different 0-dimensional ``submanifolds''
(in the algebraic geometry sense) of ${\bC}^2$. If we denote by ${\CI}_l$
the algebra of holomorphic functions (polynomials) on ${\bC}^2$ which
vanish on the D(-1) instantons, stuck to the $l$'th D3-brane, then it
can be identified with the ideal in the ring of polynomials
${\bC}[z_1, z_2]$
such that the quotient ${\bC}[z_1, z_2]/{\CI}_l$ is spanned by
the monomials
$$
z_1^{i-1} z_2^{j-1}, \qquad \qquad 1 \leq j \leq k_{li}
$$

\mezzo{Remark on Planck constant}

In what follows we set ${\e}_1 = - {\e}_2 = {\hbar}$. Note, that
this Planck constant has nothing to do with the coupling constant
of the gauge theory, where it appears as the parameter of the
geometric background \twm. It corresponds however exactly to the
loop counting in the dual string theory, while the gauge theory
Planck constant in string theory picture arises as a {\it
worldsheet} parameter, according to the relation between the
world-sheet and gauge theory instantons, described below.

\subsec{Correlation functions of the chiral operators}

Now we are
ready to attack the correlation function \gnfnc. First of all,
using the unbroken supercharges one argues that this correlation
function is independent of the coefficient in front of the term
$\vert F^{+} \vert^2 + \ldots$ which is $\{ Q, \ldots \}$.
Therefore, one can go to the weak coupling regime (with the theta
angle appropriately adjusted, so that ${\tau}_0$ is finite, while
${\bar\tau}_0 \to \infty$ ) in which \gnfnc\ is saturated by
instantons (cf. \nsvz).

In this limit the descendants of the chiral
operators become the Chern classes of the universal bundle,
``integrated'' (in the equivariant sense),  over ${\bf R}^4$. Here
is the table of equivariant integrals \DHf\ (cf. \rltns):
\eqn\tabl{ \int_{{\bf R}^4} {\Omega}^{(4)} = {{\Omega}^{(0)}(0)
\over {{\e}_1 {\e}_2}}}We should then integrate these classes over
${\CM}$. But then again, we use equivariant localization, this
time on the fixed points in ${\CM}$. These fixed points are
labelled by partitions ${\bk}$. The calculation of the expectation
values of the chiral operators becomes equivalent to the
calculation of the expectation values of some operators in the
statistical mechanical model, where the basic variables are the
$N$-tuples of partitions \prtns. In this
statistical model, the operator ${\CO}_J^{(0)} = {1\over{J}}
{\Tr}{\Phi}^J$ in the gauge theory translates to the operator
($a_l = {\hbar} M_l$): \eqn\stmop{\eqalign{& {\CO}_J [{\vec\bk}] \equiv
\left[ \int_{{\bR}^4} {\CO}^{(4)}_{J} \right]_{P_{\vec\bk}} =
{{\hbar}^{J}\over J} \times \cr & \sum_{l=1}^{N} \left[ M_l^J +
\left( \sum_{i=1}^{\infty} ( M_l + k_{li} - i +1)^J - (M_l +
k_{li} - i)^J - (M_l +  1 - i)^J + (M_l -  i )^J \right)\right]\cr
& \qquad =^{\kern -.2in \rm formally} \ {1\over J} \sum_{l, i}
\left[ \left( (a_l + {\hbar} (k_{li} + 1- i) \right)^J - \left(
a_l + {\hbar} (k_{li} - i)\right)^J \right]\cr }} This is a straightforward
consequence of \chrneva\ for ${\e}_1 = - {\e}_2 = {\hbar}$.

Given the single-trace operators ${\CO}_J$ we build arbitrary
gauge-invariant operators ${\CO}_{\vec n}$ as in \mcrsc, \vvs. After
that one can integrate their ${\CN}=2$ descendants
${\CO}_{\vec n}^{(4)}$ using the table of equivariant integrals \tabl.

Gauge theory generating function of the correlators of the chiral
operators becomes the
statistical model partition function with all the integrated operators
$\int_{{\bR}^4} {\CO}_{\vec n}^{(4)}$ added to the Hamiltonian.
In other words, we sum over the partitions
$\{{\bf k}_l\} = \{ k_{li}\}$ the Bolzmann
weights $\exp\left(-{1\over{{\hbar}^2}} \sum_{\vec n}
t_{\vec n}{\CO}_{\vec n} \right)$, and
the measure on the partitions
is given by the square of the regularized discretized Vandermonde
determinant:
\eqn\msr{\eqalign{& {\m}_{\vec\bk} = \prod_{(li) \neq (mj)}
\left( {\l}_{li} - {\l}_{mj} \right) \cr & {\l}_{li} = a_l +
{\hbar} (k_{li} - i), \quad
 \cr}}
The product in \msr\ is taken over all pairs $(li)\neq (mj)$ which is
short for $\{(l\neq m); {\rm or} \ (l=m,i\neq j); \}$ and can be
understood with the help of ${\zeta}$-regularization:
\eqn\rglmsr{{\m}_{\vec\bk} = {\exp}
\left(- {d\over ds}
{1\over{{\Gamma}(s)}} \int_{0}^{\infty} dt\ t^{s-1} \sum_{(li) \
\neq (mj)} \left.e^{-t ({\l}_{li} -
{\l}_{mj})}\right|_{s=0}\right)} The sum in \rglmsr\ is defined by
analytic continuation, as the sum over $(l,i)$ converges for
Re$({\hbar}t) < 0 $, while the sum over $(m,j)$ converges for
Re$({\hbar}t)
> 0$.

\mezzo{Remarks on literature}

At this point the reader is encouraged to consult
\nsvz\seibergone\calculus\instmeasures\twoinst\hollowood, for more conventional approach
to the instanton integrals, as well
as \maxim\torusaction\givental\atiyahsegal\adhm\DHf\ for more mathematical details.
The formula \msr\ in the case $N=2$ was shown to agree with Chern-Simons
calculations in \iqbal.

\newsec{ABELIAN THEORY}

\subsec{A little string that could}

Now suppose we take $N=1$. In the pure ${\CN}=2$ gauge theory this
is not the most interesting case, since neither perturbative, nor
non-perturbative corrections affect the low-energy prepotential.
Imagine, then, that we embed the $N=1$ ${\CN}=2$ theory in the
theory with instantons. One possibility is the \nc\ gauge theory,
another possibility is the theory on the D-brane, e.g. fractional
D3-brane at the ADE-singularity, or the D5/NS5 brane wrapping a
${\bC\bP}^1$ in {\bf K3}. In this setup the theory has
non-perturbative effects, coming from \nc\ instantons, or
fractional D(-1) branes, or the worldsheet instantons of D1
strings bound to D5, or the elementary string worldsheet
instantons in the background of NS5 brane, or an $SL_2({\bZ})$
transform thereof. In either case, we shall get the instanton
contributions to the effective prepotential. Let us calculate
them.

We shall slightly change the notation for the times ${\tau}_{\vec
n}$ as in this case there is no need to distinguish between
${\Tr}{\Phi}^J$ and $({\Tr}{\Phi})^J$. We set:
\eqn\tmss{\sum_{\vec n} {\tau}_{\vec n} \prod_{J=1}^{\infty} {x^{J
n_J} \over {{n_{J}!} (J)^{n_J}}} = \sum_{J=1}^{\infty}  t_J
{x^{J+1} \over (J+1)!} } and consider the partition function as a
function of the times $t_{J}$.

First, let us turn off the higher
order Casimirs. Then, we are to calculate:
\eqn\uonecalc{e^{-t_1
{{a^2}\over 2{\hbar}^2}} \sum_{\bk} {\m}_{\bk} e^{t_1 \vert {\bk}
\vert}}

\mezzo{Partitions and representations}

As it is well-known, the partitions ${\bk}= \left( k_{1} \geq k_{2}
\geq k_{3} \geq \ldots k_{n}\right)$ are in
one-to-one correspondence with the irreducible representations
$R_{\bk}$ of the symmetric group ${\CS}_{k}$, $k = \vert {\bk}
\vert$. Moreover, in the case $N=1$, one gets from \msr: $$
{\m}_{\bk} = \prod_{i\neq j}^\infty{\hbar(k_i-k_j+j-i)\over
\hbar(j-i)} $$ and using the relation between partitions
${\bk}$ and Young diagrams $Y_{\bk}$, whose $i$'th row contains
$k_i > 0$ boxes, $1 \leq i \leq n$ corresponding to the
irreducible representation $R_{\bk}$ of the symmetric group
${\CS}_{k}$ (and to the irreducible representation ${\CR}_{\bk}$
of the group $U({\hat N})$, for any ${\hat N} \geq n$), this can
be rewritten as
$$
{\m}_{\bk} = (-1)^{k}\left[
\prod_{i<j}^n{\left({\hbar} \left(k_i-k_j+j-i\right)\right)}
\prod_{i=1}^n{1\over {\hbar}^{k_i + n -i}
(k_i+n-i)!} \right]^2 = (-1)^k \left[ {{\rm dim}R_{\bk}
\over {\hbar}^k \ k!} \right]^2
$$
where we employ the rule ${l\cdot (l+1)\cdot (l+2)\dots \over
1\cdot 2\cdot 3\dots l\cdot (l+1)\cdot (l+2)\dots} = {1\over l!}$.
Hence \eqn\plach{ {\m}_{\bk} = \left( {{\rm dim}R_{\bk} \over k!}
\right)^2 (-{\hbar}^2)^{-k}} The measure \plach\ on the partitions
is the so-called Plancherel measure, introduced by A.M.~Vershik,
and studied extensively by himself and S.V.~Kerov \yung. Our
immediate problem is rather simple, however. The summation over
${\bk}$ is trivial thanks to Burnside's theorem, and we conclude:
\eqn\uoneansw{Z = \exp\left[ - {1\over {\hbar}^2} \left( t_1 {a^2
\over 2} + e^{t_1} \right)\right]} We see that the gauge theory
prepotential or the free energy of our statistical model coincides
with the Gromov-Witten prepotential of the ${\bC\bP}^1$
topological sigma model.


\mezzo{Back to fractional branes and to little strings}

At this point the fair question is: where this ${\bC\bP}^1$ came
from? After all, in conventional physical applications of the
topological strings the target space should be a Calabi-Yau
manifold, and ${\bC\bP}^1$ is definitely not the one. One can
imagine the topological string on a local Calabi-Yau, which is a
resolved conifold, i.e. a total space of the ${\CO}(-1) \oplus
{\CO}(-1)$ bundle over ${\bC\bP}^1$. One can then turn the
so-called twisted masses ${\m}_1, {\m}_2$, or, more mathematically
speaking, equivariant parameters with respect to the rotations of
the fiber of the vector bundle. In the limit ${\m}_{1,2} \to 0$
the sigma model is localized onto the maps into ${\bC\bP}^1$
proper. Is this the way to embed our model in a full-fledged
string compactification? We doubt it is the case.

Rather, we think the proper model should be that of little string
theory \mmm\ compactified on ${\bC\bP}^1$. Indeed, the discussion
in the beginning of this section suggests a realization of the
abelian gauge theory with instantons by means of the D5 brane
wrapping a ${\bC\bP}^1$ inside the Eguchi-Hanson space
$T^{*}{\bC\bP}^1$, which is the resolution of the
${\bC}^2/{\bZ}_2$ singularity. The wrapped D5 brane is a blown-up
fractional D3 brane stuck at the singularity. It supports an
${\CN}=2$ gauge theory with a single abelian vector multiplet. In
addition, it has instantons, coming from fractional D(-1) branes,
or, after resolution, D1 string worldsheet instantons. These are
bound to the D5 brane worldvolume. After S-duality and appropriate
decoupling limits these turn into the so-called {\it little
strings}, of which very little is known. In particular, much
debate was devoted to the issue of the tunable coupling constant
in these theories. Our results strongly suggest such a
possibility.

\subsec{Free fermions} Now let us turn on the higher order
Casimirs in the gauge theory. To facilitate the calculus it is
convenient to introduce the formalism of free fermions. Consider
the theory of a single free complex fermion on a two-sphere: $\int
{\tilde\psi} {\pb} {\psi}$. We can expand:
\eqn\frem{\eqalign{& {\psi} (z) = \sum_{r \in {\bZ} + {\half}} \
{\psi}_{r} \  z^{-r} \left( dz \over z \right)^{\half}, \cr &
{\widetilde\psi} (z) = \sum_{r \in {\bZ} + {\half}} {\widetilde
\psi}_{r} \ z^r \left( dz \over z \right)^{\half} \cr &
\qquad\qquad \{ {\psi}_{r}, {\widetilde\psi}_{s} \} = {\d}_{rs}
\cr} }The fermionic Fock space is constructed with the help of the
charge $M$ vacuum state\foot{Any $M$ is good for building the
space.}:
\eqn\vcms{\eqalign{& \vert M \rangle = {\psi}_{M + {1\over
2}} {\psi}_{M+{3\over 2}} {\psi}_{M+ {5\over 2}} \ldots \cr &
{\psi}_{r} \vert M \rangle = 0, \qquad r
> M \cr & {\widetilde\psi}_{r} \vert M \rangle = 0, \qquad r < M
\cr}} It is also convenient to use the basis of the so-called
partition states
(see, e.g. \op\kharchev). For each partition ${\bk} = ( k_1  \geq
k_2 \geq \ldots )$ we introduce the state:
\eqn\prtst{\vert M;
{\bk} \rangle = {\psi}_{M+{1\over 2} -k_1} {\psi}_{M+{3\over 2} -
k_2} \ldots }
One defines the $U(1)$ current as: \eqn\crnt{ \eqalign{& J =
: {\widetilde\psi} {\psi} : = \sum_{n \in {\bZ}} J_{n} z^{-n} {dz
\over z} \cr & J_{n} = \sum_{r < n} {\widetilde\psi}_{r}
{\psi}_{n-r}  - \sum_{r
> n} {\psi}_{n-r} {\widetilde\psi}_{r} \cr}}
Recall the bosonization rules: \eqn\bsnz{{\psi} = : e^{i {\phi}}
:\ , \quad {\tilde\psi} = : e^{-i{\phi}} :\ , \quad J =
i{\p} {\phi}} and a useful fact from $U({\hat N})$ group theory:
the famous Weyl correspondence states that
\eqn\wlcrsp{({\bC}^{{\hat N}})^{\otimes k} = \bigoplus_{{\bk}, \vert
{\bk} \vert = k} R_{\bk} \otimes {\CR}_{\bk}} as ${\CS}_{k} \times
U({\hat N})$ representation. Now let $U = {\rm diag} \left( u_1,
\ldots, u_{\hat N} \right)$ be a $U({\hat N})$ matrix. Then one
can easily show using Weyl character formula, and the standard
bosonization rules, that: \eqn\chrt{{\Tr}_{{\CR}_{\bk}} U =
\langle {\hat N}; {\bk} \vert : e^{i \sum_{n=1}^{{\hat N}} {\phi}
( u_n) } : \ : e^{- i {\hat N} {\phi} (0)}:\vert 0 \rangle \qquad }
From this formula one derives:
\eqn\prtnstts{
e^{J_{-1}\over {\hbar}} \ \vert M \rangle = \sum_{\bk}
{{\rm dim}R_{\bk} \over {\hbar}^k \ k! }\ \vert M; {\bk} \rangle}

\newsec{INTEGRABLE SYSTEM AND ${\bC\bP}^1$ SIGMA MODEL}

The importance of the
fermions is justified by the following statement.
The generating function with turned on higher Casimirs equals to
the correlation function:
\eqn\frmrp{\eqalign{& Z = {{\langle M \vert e^{J_{1} \over
{\hbar}} {\exp} \left[  \sum_{p = 1}^{\infty} {\hat t}_{p} W_{p+1}
\right] e^{-{J_{-1}\over{\hbar}}} \vert M \rangle}} \cr}}Here:
\eqn\shft{\eqalign{& \sum_{p=1}^{\infty} {\hat t}_{p} \ x^{p} =
\sum_{p=1}^{\infty} {1\over{(p+1)!}} \ t_{p} \ {{(x+{\hbar \over
2})^{p+1} - (x - {\hbar \over 2})^{p+1}}\over \hbar} \cr}}
and
\eqn\wgen{ W_{p+1}
= {1\over \hbar} \oint \ : \ {\widetilde \psi} \left( {\hbar} D
\right)^{p} {\psi} : \ , \qquad D = z{\p}_z \ }  If only $t_1
\neq 0$ the correlator \frmrp\ is trivially computed and gives
\uoneansw\ with $a=\hbar M$. From comparison of \frmrp\ with the
results of \op\ one gets that generating function \frmrp, as a
function of times $\hat t_p$, is a tau-function of the Toda lattice
hierarchy. Note that the fermionic matrix element \frmrp\ is
very much different from the standard
representation for the Toda tau-function \todalit.
In our case the ``times'' are coupled to the ``zero
modes" of higher W-generators,
while usually they couple to the components \crnt\ of the $U(1)$ current.

The free fermionic representation \frmrp\ is useful in several
respects. One of them is the remarkable mapping of the gauge
theory correlation function to the amplitudes of a (topological
type A) string, propagating on ${\bC\bP}^1$. Indeed, using the
results of \op\ (see also \prtoda)  one can show that:
\eqn\gstrcr{\biggl\langle {\exp} \int_{{\bR}^4}
\sum_{J=1}^{\infty} t_{J} \ {\CO}^{(4)}_{J+1} \biggr\rangle_{a,
{\hbar}}^{\rm gauge \ theory} = {\exp} \sum_{g=0}^{\infty}
{\hbar}^{2g-2} \langle\langle {\exp} \int_{{\Sigma}_{g}} \ a \cdot
{\bf 1} + \sum_{p=1}^{\infty} {\hat t}_{p} {\s}_{p-1}({\o})
\rangle\rangle_{g}^{\rm string}}
Here $\langle\langle\ldots\rangle\rangle_{g}$ stands for the genus $g$
connected partition function.

It is tempting to speculate that a similar relation holds for
nonabelian gauge theories. The left hand side of \gstrcr\ is known
for the gauge group $U(N)$ (we essentially described it by the
formulae \stmop\msr, see also \swi) but the right hand side is
not, although there are strong indications that the free fermion
representation and relation to the ${\bC\bP}^1$ sigma model holds
in this case too \nok.

The formula \gstrcr\ is the content of our gauge theory/string theory
correspondence. We have an explicit mapping between the gauge theory
operators and the string theory vertex operators. In this mapping the higher
Casimirs map to gravitational descendents of the K\"ahler form.

\mezzo{Full duality?}

The topological string on ${\bC\bP}^1$ actually has even more observables
then the ones presented in \gstrcr. Indeed, we are missing all the
gravitational descendants of the puncture operator ${\s}_k({\bf 1}), k> 0$.
We conjecture, that their gauge theory dual, by analogy with AdS/CFT
correspondence \bulkbndr, is the shift of vevs of the operators
${\Tr} {\phi}^J$, for ${\s}_{J-1}({\bf 1})$. For $J=1$
we are talking about shifting $a$, the vev of ${\phi}$. This is
indeed the case.
When all these couplings are taken into account we would expect
to see the full two-dimensional Toda hierarchy \todalit.

\mezzo{Chiral ring}

Another application of \frmrp\ is the
calculation of the expectation values of ${\CO}_J$. This exercise
is interesting in relation to the recent matrix model/gauge theory
correspondence of R.~Dijkgraaf and C.~Vafa \dv, which predicts,
according to \cdws: \eqn\expct{\langle {\Tr} {\phi}^J \rangle =
\oint \ x^{J} {{d z} \over z}, \qquad z + {{\Lambda}^{2N} \over z}
= P_{N} (x) = x^{N} + u_1 x^{N-1} + u_2 x^{N-2} + \ldots + u_N}
quite in
agreement with the formulae from \whitham, obtained in the context
of the Seiberg-Witten theory.

To compute the
expectation values of ${\CO}_{J}$ in our approach (for $N=1$) it
suffices to calculate $-{\hbar}^2 \
{1\over Z} {\p}_{t_{J-1}}
Z$ at $t_2 = t_3 = \ldots = 0$  and then send ${\hbar} \to 0$ (as
\cdws\ did not look at the equivariance with respect to the
space-time rotations):
\eqn\expv{\eqalign{& \langle {\CO}_{J}
\rangle_{a, 0} = \cr & = \lim_{{\hbar} \to 0}  \ {\hbar}^{J}
\ {{\langle M \vert e^{{1\over{\hbar}} \oint : {\tilde\psi} z
{\psi} : } \oint \ : {\tilde\psi} \left( ( D + {1\over 2} )^{J} -
( D - {1\over 2})^{J} \right) {\psi} : \ e^{-{{\Lambda}^2
\over\hbar} \oint : {\tilde\psi} z^{-1} {\psi} : } \vert M
\rangle}\over{\langle M \vert e^{{1\over{\hbar}} \oint :
{\tilde\psi} z {\psi} :} \ e^{-{{\Lambda}^2 \over\hbar} \oint :
{\tilde\psi} z^{-1} {\psi} : } \vert M \rangle}}= \cr
&\qquad\qquad  = \oint \left( a + z + {{\Lambda}^2 \over z}
\right)^{J} {dz \over z} \cr &  \cr & \qquad\qquad\qquad
{\Lambda}^2 = e^{t_1}, \qquad a = {\hbar} M \cr} } the last
relation proved by bosonization. This reproduces
\expct\ for $N=1$.

\newsec{THEORY WITH MATTER} In this section we
shall discuss theory with matter in the fundamental
representation. We shall again consider only $U(1)$ case, but as
above we shall be, in general, interested in turning on higher
Casimirs. To avoid the confusion, we shall use the capital letters
$T_p$ for the couplings of the theory with matter.

\subsec{4d and 2d field theory}

The famous condition of asymptotic freedom, $N_f \leq 2N_c$, if
extrapolated to the case $N_c =1$ suggests that we could add up to
two fundamental hypermultiplets. It is a straightforward exercise
to extend the fixed point calculus to incorporate the effect of
the charged matter. Let us briefly remind the important steps.
Susy equations in the presence of matter hypermultiplet $M =
({\tilde Q}, Q)$ change from $F^{+} = 0$ to $F^{+} + {\bar M}
{\Gamma} M = 0$, ${\dir} M = 0$. The moduli space of solutions to
these equations looks near $M=0$ locus as  a vector bundle over
${\CM}$ -- the instanton moduli, whose fiber is the bundle of
Dirac zero modes.

It can be shown that the instanton measure gets an extra factor,
the equivariant Euler class of the Dirac bundle (see \issues\ for
more details and more references). The localization formulae
still work, but now each partition ${\bk}$ has an extra weight
\swi. The contribution of the fixed point to the path integral in
the presence of the matter fields is \msr\ multiplied by the extra
factor (the content polynomial \macdonald):
\eqn\extrw{{\tilde\m}_{\bk} (a,m) = Z^{pert} (a, m) \times
\prod_{f=1}^2 \prod_{i=1}^{\infty} \left( a + m_f + {\hbar}( 1 - i
) \right) \ldots\left( a + m_f + {\hbar} (k_i - i) \right)} where
\eqn\zper{\eqalign{& Z^{pert} (a, m) = \prod_{f}
\prod_{i=1}^{\infty} {\Gamma}\left( {a + m_f \over {\hbar}}  + 1-
i \right) \sim {\exp} \int_{0}^{\infty} {{\rm d}t \over t} \sum_f
{e^{- t ( a + m_f )} \over {\rm sinh}^2 \left( {{\hbar} t \over 2
} \right)} = \cr & = \sum_{f} \left[ {( a + m_f)^2 \over
2{\hbar}^2}  {\rm log} (a+m_f) + {1\over 12} {\rm log} ( a + m_f)
+ \sum_{g>1} {B_{2g}\over 2g(2g-2)} \left( {{\hbar}\over{a+m_f}}
\right)^{2g-2} \right]\cr}}
The bosonization rule \chrt\ leads to
the following formula:
\eqn\instcntr{Z^{inst} = \biggl\langle e^{i {a + m_2 \over {\hbar}
} {\phi} ({\infty}) } e^{-i {m_2 \over {\hbar}} {\phi} (1) }
e^{\sum_p T_p {W}_{p+1} } e^{i {m_1 \over {\hbar}} {\phi}(1)}
e^{ - i {{a+m_1}\over{\hbar}} {\phi}(0)} \biggr\rangle}
It can be shown that the full partition function $Z^{pert}
Z^{inst}$ also has a CFT interpretation, and also obeys Toda
lattice equations. We shall discuss this in a future publication.

\subsec{Relation to geometric engineering}

Now let us turn off the higher Casimirs, i.e. set $T_p = 0$, for
$p
> 1$ . Then \instcntr\extrw\ lead to \eqn\prptns{\eqalign{&
{\CF}_{0} = {\half}T_1 a^2 - m_1 m_2 {\rm log} ( 1 - e^{T_1} )+
\sum_f {\half} \left(a+m_f\right)^2 {\rm log} \left( a + m_f
\right) \cr & {\CF}_1 = {1\over 12} {\rm log} ( a+m_1)( a+m_2) \cr
& {\CF}_{g} = {B_{2g} \over 2g (2g-2) } \sum_f {{\hbar}^{2g-2}
\over{( a+m_f)^{2g-2}}} \cr}}

We remark that \prptns\ is a limit of the all-genus topological
string prepotential in the geometry described in \agmav\ (Fig.12,
Eq. (7.34)). The specific limit is to take first $t_1, t_2, g_s$
in their notation to zero, as $t_f = {\b} ( a + m_f), g_s = {\b}
{\hbar}$, ${\b} \to 0$, while $-r^{\prime}$ (their notation) $=
T_1$ (our notation) is finite. The prepotential \agmav\ actually
describes the five dimensional susy gauge theory compactified on a
circle of circumference ${\b}$. The limit ${\b} \to 0$ actually
takes us to the four dimensional theory, which is what we were
studying in this paper. It is clear, from \agmav\ (Fig.12c) that
the geometry corresponds to the $U(1)$ gauge theory with two
fundamental hypermultiplets (two D-branes pulling on the sides).

Our results are, however, stronger. Indeed, we were able to
calculate the prepotential and ${\CF}_g$'s with arbitrary higher
Casimirs turned on. In the limit \eqn\lmts{m_1, m_2 \to \infty,
\quad e^{T_1} \to 0, \quad{\rm  so \ that} \quad {\Lambda}^2 = m_1
m_2 e^{T_1}= e^{t_1} \quad {\rm  our \ old \ notation, \ is \
finite}} we get back the pure $U(1)$ theory, which we identified
with the topological string on ${\bC\bP}^1$ \gstrcr. Note that
this was not ${\bC\bP}^1$ embedded into Calabi-Yau, as in the
latter case no gravitational descendants ever showed up. We are
led, therefore, to the conclusion, that the topological string on
the geometry of Fig.12 of \agmav\ has a deformation, allowing
gravitational descendants, and flowing, in the limit \lmts\ to the
pure ${\bC\bP}^1$ model. This fascinating prediction certainly
deserves further study.

\bigskip
\noindent {\it
Acknowledgements.}

NN acknowledges useful discussions with N.~Berkovitz, S.~Cherkis,
M.~Douglas,  A.~Givental, D.~Gross, M.~Kontsevich, G.~Moore,
A.~Polyakov, N.~Seiberg, S.~Shatashvili, E.~Witten, C.~Vafa, and
especially A.~Okounkov. We also thank A.~Gorsky, S.~Kharchev,
A.~Mironov, A.~Orlov, V.~Roubtsov, S.~Theisen and A.~Zabrodin for
their help. NN is grateful to Rutgers University, Institute for
Advanced Study, Kavli Institute for Theoretical Physics, and Clay
Mathematical Institute for support and hospitality during the
preparation of the manuscript. ASL and AM are grateful to IHES for
hospitality, AM acknowledges the support of the Ecole Normale
Superieure, CNRS and the Max Planck Institute for Mathematics
where this work was completed. Research was partially supported by
{\cyr RFFI} grants 01-01-00548 (ASL),
02-02-16496 (AM) and
01-01-00549 (NN) and by the INTAS grant 99-590 (ASL and AM).

\appendix{A}{Equivariant integration and localization}

Let ${\CY}$ be a  manifold with an action of a Lie group $\CG$,
and let $\CX$ be a $\CG$-invariant submanifold. Moreover, let
$\CX$ be a zero locus of a section $s$ of a $\CG$-equivariant
vector bundle ${\CV}$ over $\CY$.

Suppose that we need to develop an integration theory on the
quotient ${\CX}/\CG$. It is sometimes convenient to work
$\CG$-equivariantly on $\CY$, and use the so-called Mathai-Quillen
representative of the Euler class of the bundle ${\CV}$.

The equivariant cohomology classes are represented with the help
of the equivariant forms. These are functions on ${\bf g} = Lie({\CG})$
with the
values in the de Rham complex of ${\CY}$. In addition, these
functions are required to be $\CG$-equivariant, i.e. the adjoint
action of $\CG$ on $\bf g$ must commute with the action of $\CG$
on the differential forms on $\CY$.

Let us denote the local coordinates on $\CY$ by $y^{\m}$, and
their exterior differentials $dy^{\m}$ by $\psi^{\m}$. The
equivariant differential is the operator \eqn\eqvo{Q = {\psi}^{\m}
{{\p}\over{{\p}y^{\m}}} + {\phi}^a V_a^{\m}
(y){{\p}\over{{\p}{\psi}^{\m}}}} where ${\phi}^a$ are the linear
coordinates on $\bf g$, and $V_a = V^{\m}_a {\p}_{\m}$ are the
vector fields on $\CY$ generating the action of $\CG$. The
operator $Q$ raises the so-called ghost number by one: \eqn\ghi{
gh = {\psi} {{\p}\over{{\p}{\psi}}} + 2 {\phi}
{{\p}\over{{\p\phi}}} } The equivariant differential forms can be
now written as $\CG$-invariant functions of $(y,\psi,\phi)$.

In the applications one uses a more refined (Dolbeault) version of
the equivariant cohomology. There, one multiplies ${\CY}$ by $\bf
g$, and extends the action of $\CG$ by the adjoint action on $\bf
g$. The coordinate on this copy of $\bf g$ is conventionally
denoted by $\bar\phi$, and its differential by $\eta$. The
equivariant differential on $\CY \times \bf g$ acts, obviously,
as: \eqn\eqvii{Q = {\psi}^{\m} {{\p}\over{{\p}y^{\m}}} + {\phi}^a
V_a^{\m} (y){{\p}\over{{\p}{\psi}^{\m}}} + {\eta}
{{\p}\over{\p\bar\phi}} + [{\phi}, {\bar\phi}]
{{\p}\over\p\eta}}However, the  ghost number is defined not as in
\ghi\ but rather with a shift (in some papers this shift is
reflected by the notation ${\bf g}[-2]$): \eqn\ghii{gh = {\psi}
{{\p}\over{{\p}{\psi}}} + 2 {\phi} {{\p}\over{{\p\phi}}} -
2{\bar\phi}{{\p}\over{\p\bar\phi}} - {\eta}{{\p}\over\p\eta}}
Suppose ${\CO} ( y, {\psi}, {\bar\phi}, {\eta}, {\phi})$ is $\bG$
invariant and annihilated by $Q$. Suppose in addition that the
following integral makes sense:
\eqn\eqvint{{\CI}_{{\CO}}({\bar\phi}, \phi, \eta) = \int dy d\psi
\ {\CO}(y,\psi, {\bar\phi}, \phi, \eta)} Then ${\CI}_{\CO}$ is
$\bG$-equivariant on $\bf g$. One can integrate it over $\bar\phi,
\eta$, and $\phi$ against any $\CG$ equivariant form.

\mezzo{Amplitudes}

In particular, one can simply integrate ${\CI}_{\CO}$ over all of
$\bf g$: \eqn\fldsn{{\CI}_{\CO}^{top} = \int d{\bar\phi}d{\eta}
{{d{\phi}}\over{{\rm Vol}({\CG})}} \ {\CI}_{\CO} ({\bar\phi},
\phi, \eta)}

More general construction proceeds by picking a normal subgroup $H
\subset {\CG}$, and integrating over $Lie ({\CG}/H)$, with an
extra measure: \eqn\prtgf{{\CI}^{H}_{\CO} ( {\vf} )_{\kappa} =
\int {{d{\bar\phi}^{\perp} d{\eta}^{\perp}
d{\phi}^{\perp}}\over{{\rm Vol}({\CG}/H)}} \
{\CI}_{\CO}({\bar\phi}^{\perp} + {\bar\vf}, {\eta}^{\perp},
{\phi}^{\perp} +{\vf}) \ e^{-{1\over \kappa} \left( \Vert [
{\bar\phi}^{\perp} + {\bar\vf} , {\phi}^{\perp} + {\vf} ] \Vert^2
- \langle [{\eta}^{\perp}, {\phi}^{\perp} + {\vf}],
{\eta}^{\perp}\rangle\right)}} where ${\phi}^{\perp} \in
Lie({\CG}/H)$ etc., ${\vf},{\bar\vf} \in Lie(H)$, and as long as
$[{\vf}, {\bar\vf}]= 0$ the left hand side of
${\CI}^{H}_{\CO}({\vf})$ does not depend on ${\bar\vf}$, as a
consequence of $Q$-symmetry. Clearly ${\CI}^{top} =
\left.{\CI}^{1}\right|_{\kappa=\infty}$.

Now let us sophisticate our construction a little bit more. Recall
that we had a vector bundle $\CV$ over ${\CY}$, with the section
$s = (s^a(y))$. Suppose, in addition, that there is a
$\bG$-invariant metric $g_{ab}$ on the fibers of $\CV$, and let
${\Gamma}_{{\m} a}^b dy^{\m}$ denote a connection on $\CV$,
compatible with $g_{ab}$. Then the following integral produces a
$Q$-invariant form: \eqn\elrcs{{\CO}_{\CV}(y,\psi, {\phi},
{\bar\phi}, {\eta}) = \int d{\chi}_a dH_a \ e^{i {\chi}_a
{\psi}^{\m} \left( {\p}_{\m} + {\Gamma}_{\m} \right)s^a + i H_a
s^a - {\half} g^{ab} \left[ H_a H_b + \left( {\CF}_{\m\n, a}^{c}
{\psi}^{\m}{\psi}^{\n} + R({\phi},y)^{c}_{a}  \right) {\chi}_c
{\chi}_b \right]}} where ${\CF} = d{\Gamma} + [{\Gamma},{\Gamma}]$
is the curvature of ${\Gamma}$, and $R({\phi},y)$ is the
representation of $\bf g$, acting on the sections of $\CV$ (a Lie
algebraic 1-cocycle).

Now, if we rescale the metric $g^{ab} \to t g^{ab}$ then the value of
\elrcs\ should not change (the variation is $Q$-exact). In particular, in
the limit $t \to 0$ the form ${\CO}_{\CV}$ is supported on the zeroes of the
section $s$. In the opposite limit, $t \to \infty$ it becomes independent of
$s$ and turns into a form:
$$
{\CO}_{\CV} \sim {\rm Pf} \left( {\CF} + R({\phi},y) \right)
$$
One can also consider more general variations of the metric $g^{ab}$.

\mezzo{Localization}

Let us go back to \prtgf. As we said, the answer is independent of
${\bar\vf}$. Let us make a good use of this fact. To this end, let us
multiply
${\CO}$ in \eqvint\ by an extra factor:
$$
{\CO} e^{-Q ( {\bar\phi}^a V^{\m}_{a} g_{\m\n} )}
$$
where $g_{\m\n}$ is any ${\CG}$-invariant metric on ${\CY}$.
Explicitly, we have got in the exponential
$$
g(V_a, V_b) {\phi}^b {\bar\phi}^a + {\rm fermions}
$$
Now let us take the limit ${\bar\vf} \to \infty$. The measure will be
localized near the zeroes of the vector field
$V_a {\bar\vf}^a$. This is the source of equivariant localization.
Say, take $H = {\CG}$ (more general case can be easily worked out). Then:
\eqn\loci{{\CI}^{\CG}_{\k}({\vf}) =
\sum_{p \in F} {{\CO}(p, {\vf}) \over \prod_i w_i({\vf})}}
where:
$F$ is the set of points on $\CY$ where $V_a {\vf}^a$ vanishes, $w_i({\vf})$
are the weights of the action of $\CG$ on the tangent space to ${\CY}$ at $p$.

\footatend\vfill\supereject\immediate\closeout\rfile\writestoppt
\baselineskip=14pt\centerline{{\bf References}}\bigskip{\frenchspacing%
\parindent=20pt\escapechar=` \input refs.tmp\vfill\eject}\nonfrenchspacing\writetoc\bye